\documentclass[useAMS,usenatbib]{mnras}
\usepackage[latin1]{inputenc}
\usepackage{verbatim}
\usepackage{macros}
\usepackage{graphicx}
\usepackage{amsmath}
\usepackage[usenames]{xcolor}
\usepackage{footmisc}
\usepackage{hyperref}
\usepackage{float}
\title
[SELGIFS data challenge]
{The SELGIFS data challenge: generating synthetic observations of CALIFA 
galaxies from hydrodynamical simulations}

\author[G.~Guidi et al.]
{
G.~Guidi$^{1}$\thanks{E-mail:gguidi@aip.de}, J.~Casado$^{1,2,3}$, Y.~Ascasibar$^{2,3}$, R. Garc\'{i}a-Benito$^{4}$, L.~Galbany$^{5}$, 
\\~\\
{\normalfont\LARGE
P.~S\'anchez-Bl\'azquez$^{2,3,6}$, S.~F.~S\'anchez$^{7}$, F. F. Rosales-Ortega$^{8}$ and C. Scannapieco$^{1,9}$
}
\\~\\
$^{1}$ Leibniz-Institut f{\"u}r Astrophysik Potsdam (AIP), D-14482, Potsdam, Germany\\
$^{2}$ Universidad Aut\'onoma de Madrid, 28049 Madrid, Spain\\
$^{3}$ Astro-UAM, UAM, Unidad Asociada CSIC\\
$^{4}$ Instituto de Astrof\'{i}sica de Andaluc\'{i}a (IAA/CSIC), 18080 Granada, Spain\\
$^{5}$PITT PACC, Department of Physics and Astronomy, University of Pittsburgh, Pittsburgh, PA 15260, USA.\\
$^{6}$Instituto de Astrof\'isica, Pontificia Universidad Cat\'olica de
Chile, Av. Vicu\~na Mackenna 4860, 7820436 Macul, Santiago, Chile \\
$^{7}$Instituto de Astronom\'{i}a, Universidad Nacional Aut\'onoma de M\'exico, A.P. 70-264, 04519, M\'exico\\
$^{8}$Instituto Nacional de Astrof\'isica, \'Optica y Electr\'onica (INAOE), 72840 Tonantzintla, Puebla, M\'exico\\
$^{9}$Departamento de F\'isica, Universidad de Buenos Aires, Ciudad 
Universitaria, Pabell\'on I, 1428 Buenos Aires, Argentina
}

\begin{document}

\date{Accepted \today \ Received ...; in original form ...}

\pagerange{\pageref{firstpage}--\pageref{lastpage}} \pubyear{2015}

\maketitle
\label{firstpage}

\begin{abstract}
In this work we present a set of synthetic observations that mimic 
the properties of the Integral Field Spectroscopy (IFS) survey CALIFA, 
generated using radiative transfer techniques applied to hydrodynamical 
simulations of galaxies in a cosmological context. The simulated 
spatially-resolved spectra include stellar and nebular emission, 
kinematic broadening of the lines, and dust extinction and scattering.
The results of the radiative transfer simulations have been post-processed
to reproduce the main properties of the CALIFA 
V500 and V1200 observational setups.
The data has been further formatted to mimic the CALIFA survey in terms of 
field of view size, spectral range and sampling. 
We have included the effect of the spatial and 
spectral Point Spread Functions affecting CALIFA observations, and 
added detector noise after characterizing it 
on a sample of 367 galaxies.
The simulated datacubes are suited to be analysed by the same 
algorithms used on real IFS data. 
In order to provide a benchmark to compare the results obtained applying 
IFS observational techniques to our synthetic datacubes, and test the 
calibration and accuracy of the analysis tools, we have computed the 
spatially-resolved properties of the simulations.
Hence, we provide maps derived directly from the 
hydrodynamical snapshots or the noiseless spectra, in a way that is 
consistent with the values recovered by the observational analysis 
algorithms. Both the synthetic observations and the product datacubes
are public and can be found in the collaboration website \url{http://astro.ft.uam.es/selgifs/data_challenge/}.
\end{abstract}


%--------------------------------------------------------------------------
\section{Introduction}
%--------------------------------------------------------------------------

Over the last two decades, Integral-Field Spectroscopy (IFS) has become 
a standard technique to study galaxy formation and evolution over cosmic time.
Compared to single-fibre or long-slit spectroscopy, IFS allows to 
simultaneously recover the full spatial and spectral information of the 
target object.
The optical spectrum of a galaxy, or a part thereof, comprises information 
about the different components that emit or absorb light within the observed 
region, and therefore spatially-resolved spectroscopy over a significant 
extent of a galaxy provides an unprecedented level of detail on the local 
physical properties of its gas, dust, and stars, as well as valuable 
constraints on other important variables, such as its dark matter content, 
or the evolutionary path that the system may have followed to reach its 
state at the time of observation.

Nowadays, several observational programmes have produced, or will soon 
provide, systematic IFS surveys targeted at different galaxy populations, 
both in the local Universe, such as e.g. SAURON \citep{Bacon01}, 
DiskMass \citep{Bershady10}, PINGs \citep{Rosales-Ortega10}, 
ATLAS$^{\text{3D}}$ \citep{Cappellari11}, 
CALIFA \citep{Sanchez12}, SAMI \citep{Croom12}, MaNGA \citep{Bundy15}, 
MUSE \citep{Bacon04} or AMUSING \citep{Galbany16_AMUSING}, 
as well as at high redshift, such as e.g. SINS \citep{Foerster-Schreiber09}, 
KMOS$^\text{3D}$ \citep{Wisnioski15}, or KROSS \citep{Stott16}.
Although all these datasets differ widely in terms of both the number 
of galaxies observed and the number of spaxels sampling each object, 
the total number of spectra is, in most cases, so large that a 
significant part of the analysis must necessarily rely on fully automated 
procedures.
In the near future, instruments such as 
WEAVE \citep{Dalton14}, HARMONI \citep{Thatte14}, 
MIRI \citep{Wright08} and NIRSpec \citep{Birkmann10}
will routinely 
produce even larger datasets just for a single galaxy, and their likely 
use in survey mode will increase the number of spectra to be analysed by 
several orders of magnitude.

Albeit spectroscopic data allow in principle to infer the physical properties 
of the observed galaxies at a high level of detail, 
the correctness of the determination 
strongly relies on the accuracy of the different tools 
and procedures applied in 
the analysis of the spectra. Hence, the calibration of the analysis procedures, 
together 
with a rigorous assessment of the associated biases, uncertainties, model 
dependencies and degeneracies, is of paramount importance.
Many of the tools developed in the context of traditional spectroscopy, 
often aimed at disentangling the emission of gas and stars, determining 
the kinematics of either/both components, and/or reconstructing the star 
formation history by means of stellar population synthesis, usually include a 
discussion of this kind of issues in the description of their methodology
\citep[e.g.][]{Cappellari04, Cid_Fernandes05, Ocvirk06, Sarzi06, 
Koleva09, MacArthur09, Walcher11, Walcher15, Sanchez16}.

Compared to traditional spectroscopy, the IFS technique provides much more 
information about the observed
galaxies,  at the price of 
a higher level of complexity in the analysis of the data.
In particular, the precise way in which spatial information is treated 
may have a crucial impact on the feasibility of any scientific case as 
a function of the signal-to-noise ratio ($S/N$) of the observations.
Ideally, one would like to take advantage of the highest spatial 
resolution provided by the instrument, analysing every spaxel as an 
independent spectrum, but $S/N$ quickly decreases to potentially 
unacceptable levels as the incoming light is divided into many 
wavelengths and spaxels.
In order to find a trade-off between spatial resolution and $S/N$, 
several algorithms have been developed to carry out a spatial 
segmentation (binning) of the IFS datacubes based on a variety of 
different approaches \citep[see e.g.][]{Stetson87, Bertin96, 
Sanders01, Papaderos02, Cappellari03, Diehl06, 
Sanchez12_HIIexplorer, Sanchez16, Casado16}.
In fact, one of the advantages of IFS over traditional spectroscopy 
is that large areas may be combined in order to properly characterize 
weak signals.
As pointed out by e.g. \citet{Casado16}, the optimal strategy for the 
segmentation is completely dependent on the specific problem under 
consideration, and a thorough study is necessary on a case-by-case basis.

One possible approach to test the analysis tools and methodology used 
both with traditional and IFS data is to apply them on simulated spectra 
created from analytical models of the galaxies' stellar and dust content, 
or from cosmological hydrodynamical 
simulations of galaxy formation. 
The main advantage of this kind of experiments 
with respect to a purely observational approach 
is that the \emph{correct} solution to be recovered (i.e. 
the physical properties of the galaxies) is accurately 
known, which makes possible to detect, quantify, and perhaps even 
correct systematic errors.
Since current hydrodynamical codes (\citealt{Governato10, Aumer13, 
Vogelsberger14, Wang15, Governato07, Scannapieco08, Nelson15, Schaye15}
among many others) self-consistently follow the intertwined 
evolution of gas, dark matter and stars over cosmic time 
implementing a significant part of the relevant physics at the 
sub-resolution level through simple numerical schemes
(whose details have a significant influence on the results, 
see e.g. \citealt{Scannapieco12}),
they are able to
connect the observable properties of the galaxies with their merger 
and accretion history, providing useful initial
conditions for the creation of simulated spectra, with a 
complexity similar to the one of real galaxies.
As pointed out in \citet{Guidi15}, 
several uncertainties both in the hydrodynamical codes 
and in the techniques used to simulate the spectra still exist, 
and scientists must be aware of these limitations when 
conducting such studies.

When simulations are compared with observational data
of one particular instrument/survey, after creating 
the spectra of the simulations a crucial point is   
to generate a full `synthetic observation', mimicking as closely 
as possible all the known 
selection effects and biases inherent to the particular 
instrument, and then processing these data with the same 
algorithms and techniques that are applied to the actual observations, 
as done by e.g.
\citet{Scannapieco10, Belovary14, Michalowski14, Hayward14, 
Smith15, Hayward15, Guidi15, Guidi16}.
Recently, some efforts in producing mock data in the context of IFS, 
modelling the galaxy spectra using simple recipes for the stellar 
population content have been undertaken by 
\citet{Kendrew16}, who have used the {\sc hsim} pipeline 
\citep{Zieleniewski15} to create synthetic observations of simulated 
high-redshift galaxies that reproduce the conditions of the HARMONI 
instrument \citep{Thatte14}, to test its capabilities in recovering 
the stellar kinematics.

In this work we have developed a pipeline to generate IFS synthetic 
observations mimicking the Calar Alto Legacy Integral-Field Area (CALIFA) 
survey \citep{Sanchez12} from hydrodynamical simulations of galaxies. 
{A similar approach was adopted in \citet{Wild14} to analyse the major merger in NGC4676 \citep[the `Mice' galaxy pair;][]{Voronstov-Vel'Iaminov58}, using tailored simulations based on several simplifying assumptions (e.g. NFW-like profile for the dark matter halo, stellar populations exponential distribution with and without a bulge component, smooth star formation history, no chemical evolution or dust...) to study the origin
of the observed morphological and kinematic features of the merger.
Here we extend and generalize this methodology in order to provide the
community with a set of realistic virtual observations.
On the one hand, we use state-of-the-art cosmological simulations of galaxies
that take into account the most relevant
physical processes associated to galaxy formation
\citep{Scannapieco05, Scannapieco06, Aumer13}, and we use the radiative
transfer code {\sc sunrise} \citep{Jonsson06, Jonsson10} to calculate
the spatially-resolved Spectral Energy Distributions (SEDs) of our
simulated objects, fully-consistently treating
the transfer of light in the dusty ISM. On the other
hand, we model the most important observational effects of the PMAS/PPak
instrument, such as detector noise, spatial and spectral Point Spread
Functions (PSFs), as well as the three-pointing dithering and interpolation
strategy applied in the CALIFA survey.
Our final products, publicly available through a web 
interface\footnote{\url{http://astro.ft.uam.es/selgifs/data_challenge/}}, 
consist of CALIFA-like synthetic datacubes, as well as resolved maps of
the intrinsic SED (free from instrumental effects),
measurements of observable quantities such as emission-line intensities
and absorption-line indices,
and physical properties (e.g. stellar mass, age, or metallicity) of the simulated galaxies.

It is one of the long-term goals of the SELGIFS collaboration
to use the proposed `Data Challenge' to 
carefully evaluate the merits and drawbacks of different strategies that 
may be followed in order to infer these physical properties from real IFS
data.
This dataset makes possible to disentangle the well-known degeneracy
  between instrumental and methodological effects inherent to the analysis
  of astronomical observations.
  Instrumental uncertainties (e.g. noise, PSF) affect the reconstruction of the
  SED from the data, whereas the methodological aspects (e.g. explicit
  or implicit assumptions/approximations) are more relevant to the physical
  interpretation and the recovery of derived properties
%  (as opposed to the \emph{measured} SED)
  of the object.
  In particular, our synthetic observations, together with the "correct"
  solutions for the optical SED and some of the most widely studied physical
  properties of the galaxies, is ideally suited to test common assumptions,
  such as e.g. uncorrelated measurements and noise (an instrumental effect)
  or uniform-screen dust extinction (a methodological approximation),
  quantifying the magnitude of the associated systematic errors (biases
  with respect to the true solution) and the accuracy of the estimated errors.
  Tracking the reasons behind the observed discrepancies offers an
  opportunity to gain a better understanding of the different techniques,
  identify their weakest points, and hopefully devise a way to overcome them.

The structure of the paper is as follows. We illustrate in 
Section~\ref{sec:simulations} the set of hydrodynamical simulations used 
in this project, and we present the calculation of their { intrinsic
  physical}
properties. In Section~\ref{sec:simulated_spectra} we describe the procedure 
followed to generate the spectra of the simulations using the radiative transfer
code, and we explain how we calculate some of the { most widely studied
  observable}
properties from the simulated spectra.
In Section~\ref{sec:califa} we illustrate the main features of the 
CALIFA survey, as well as the technical properties { that are}
reproduced in our synthetic datacubes.
%We present the data format of the simulated dataset and of the 
%resolved maps in Section~\ref{sec:selgifs}, and 
We briefly present a simple example of the science that is enabled by this data set in Section~\ref{sec:science}, and we summarize our work in Section~\ref{sec:summary}.

%--------------------------------------------------------------------------
\section{Hydrodynamical simulations}
\label{sec:simulations}
%--------------------------------------------------------------------------

\begin{table*}
\begin{center}
\begin{tabular}{ccccccc}
\hline
Name & Total mass & Stellar age (log [yr]) & Stellar metallicity (log [$Z/Z_{\odot}$]) & v$_{\text{disp}}$ & Gas metallicity \\
& $\log (M_*/M_{\odot})$ & $\log \langle age \rangle_M$ \hspace{0.03cm} $\log \langle age \rangle_L$ & $\log \langle Z \rangle_M$ \hspace{0.3cm} $\log \langle Z \rangle_L$ & [km/s] & [12+log (O/H)] \\
\hline
C-CS$^+$ & 10.66 & 10.01 \hspace{1cm} 9.93 & -0.39 \hspace{1cm} -0.37 & 92.8 & 8.52 \\
E-CS$^+$ & 10.21  & 10.00 \hspace{1cm} 9.91 & -0.44 \hspace{1cm} -0.49 & 61.2 & 8.24 \\
D-MA & 10.75 &  9.84 \hspace{1cm} 9.68 & -0.19 \hspace{1cm} -0.05 & 63.3 & 9.09 \\
\hline
\end{tabular}
\end{center}
\caption{Global properties of the simulated galaxies used to generate the 
mock datacubes. These properties have been calculated in a 60 
kpc$\times$60 kpc region with face-on orientation. Edge-on values 
differ slightly from the ones presented here, and can be found in 
\citet{Guidi15} together with several other physical properties, while in \citet{Guidi16}
these galaxies have been compared with the Sloan Digital Sky Survey dataset 
\citep{Abazajian09}.
}
\label{tab:galaxy_properties}
\end{table*}

To produce our mock data sample we use three 
hydrodynamical simulations of galaxies in a 
$\Lambda$CDM Universe, generated with the zoom-in technique 
\citep{Tormen97} using as initial conditions 
three dark-matter halos of the Aquarius simulation \citep{Springel08}. 
The galaxies are similar to the Milky Way 
in mass ($M_{\rm vir}\sim 0.7 - 1.7\times 10^{12}$~M$_\odot$) and 
merger history \citep[see][]{Scannapieco09}.
The mass resolution is 
$1-2\times 10^{6}$~M$_\odot$ for dark matter particles and 
$2-5\times 10^{5}$~M$_\odot$ for 
stellar/gas particles. The gravitational softening at redshift zero is $300-700$~pc.

All simulations are based on the Tree-PM SPH Gadget-3 code 
\citep{Springel05} with the additional implementation of sub-resolution
numerical schemes to describe
gas cooling \citep{Sutherland93, Wiersma09},
a multi-phase InterStellar Medium (ISM) \citep{Scannapieco06},
star formation \citep{SH03},
chemical enrichment from SNe \citep{Portinari98, Chieffi04}
and AGB stars \citep{Portinari98, Marigo01, Karakas10},
as well as SNe feedback \citep{Scannapieco06, Aumer13}.
The C-CS$^+$ and E-CS$^+$ galaxies are generated with the code described in  
\citet{Scannapieco05, Scannapieco06} with updated metal yields, 
while the D-MA object has been simulated with the \citet{Aumer13} feedback model.
These simulations belong to a larger set that has already been 
studied in several works \citep[e.g.][]{Aumer13, Guidi15, 
Guidi16}, and we refer the reader to those studies for more details.

We compute the following global properties
(listed in Table~\ref{tab:galaxy_properties})
by considering all particles belonging to the main halo within a 
60 kpc$\times$60 kpc region centred around the galaxy, 
oriented face-on according to the direction of the total angular momentum
\citep{Guidi15, Guidi16}:

\begin{itemize}

\item Total stellar mass %$\log (M_* / M_\odot)$.

\item Mean stellar age / metallicity:
calculated weighting\footnote{Notice 
that in this work we use arithmetic means both for the ages and metallicities \citep{Asari07, Cid_Fernandes13}. 
A different definition often found in the literature is the geometric mean
$\langle \log X \rangle_{M} = \frac{\sum_i  M_i \cdot \log X_i }{\sum_i M_i}$
(e.g. \citealt{Gallazzi05, Gonzalez_Delgado14, Gonzalez_Delgado15, 
Sanchez16_FIT3D}). Since we will provide these quantities 
smoothed by the CALIFA spatial PSF (Sec.~\ref{sec:product_datacubes}), we choose to 
weight the linear quantities in order to avoid biases in the calculation of the
smoothed properties.}
stellar particles by their mass $M_i$
\begin{equation}
\log \langle X \rangle_M = \log \left[ \frac{\sum_i  M_i \cdot X_i }{\sum_i M_i} \right]
 \label{eq:age_mass_w}
\end{equation}
and by their luminosity $L_i$ at 5635 \AA\ 
according to the {\sc starburst 99} SPS model, for consistency with
previous observational studies of CALIFA galaxies \citep[e.g][]{Gonzalez_Delgado14, Sanchez16, Ruiz_Lara16}
%in the $r$-band $L_i$
%according to the \citet{Bruzual03} SPS mode
\begin{equation}
 \log \langle X \rangle_L = \log \left[ \frac{\sum_i  L_i \cdot X_i }{\sum_i L_i} \right]
 \label{eq:age_lum_w}
\end{equation}

\item Luminosity-weighted velocity dispersion in the face-on projection:
%\begin{eqnarray}
%V_{\text{disp}} &=& \sqrt{\frac{1}{N-1}\sum_{i=1}^{N} (v_i - \bar{V})^2}
%\end{eqnarray}
%where $\bar{V}$ is the mean velocity.
\begin{equation}
  v_{\text{disp}} = \sqrt{\frac{\sum_i L_i \cdot (v_i - \bar{v})^2}{\sum_i L_i}}
   \label{eq:v_disp}
\end{equation}
where $\bar{v}$ is the mean velocity
\begin{equation}
  \bar{v} = \frac{\sum_i  L_i \cdot v_i }{\sum_i L_i} \\
   \label{eq:v_mean}
\end{equation}
%Units are km~s$^{-1}$.

\item Mass-weighted gas metallicity:
\begin{equation}
12 + \log(O/H) = 12 + \log \left[\frac{ \sum_{\rm gas} M_i \cdot (O/H)_i }{ \sum_{\rm gas} M_i }\right]
\end{equation}

\end{itemize}

%--------------------------------------------------------------------------
% Resolved properties

%%%%%%%%%%%%%%%%%%%%%%%%%%%%%
%\begin{comment}

\begin{figure}
\centerline{\includegraphics[width=.43\textwidth]{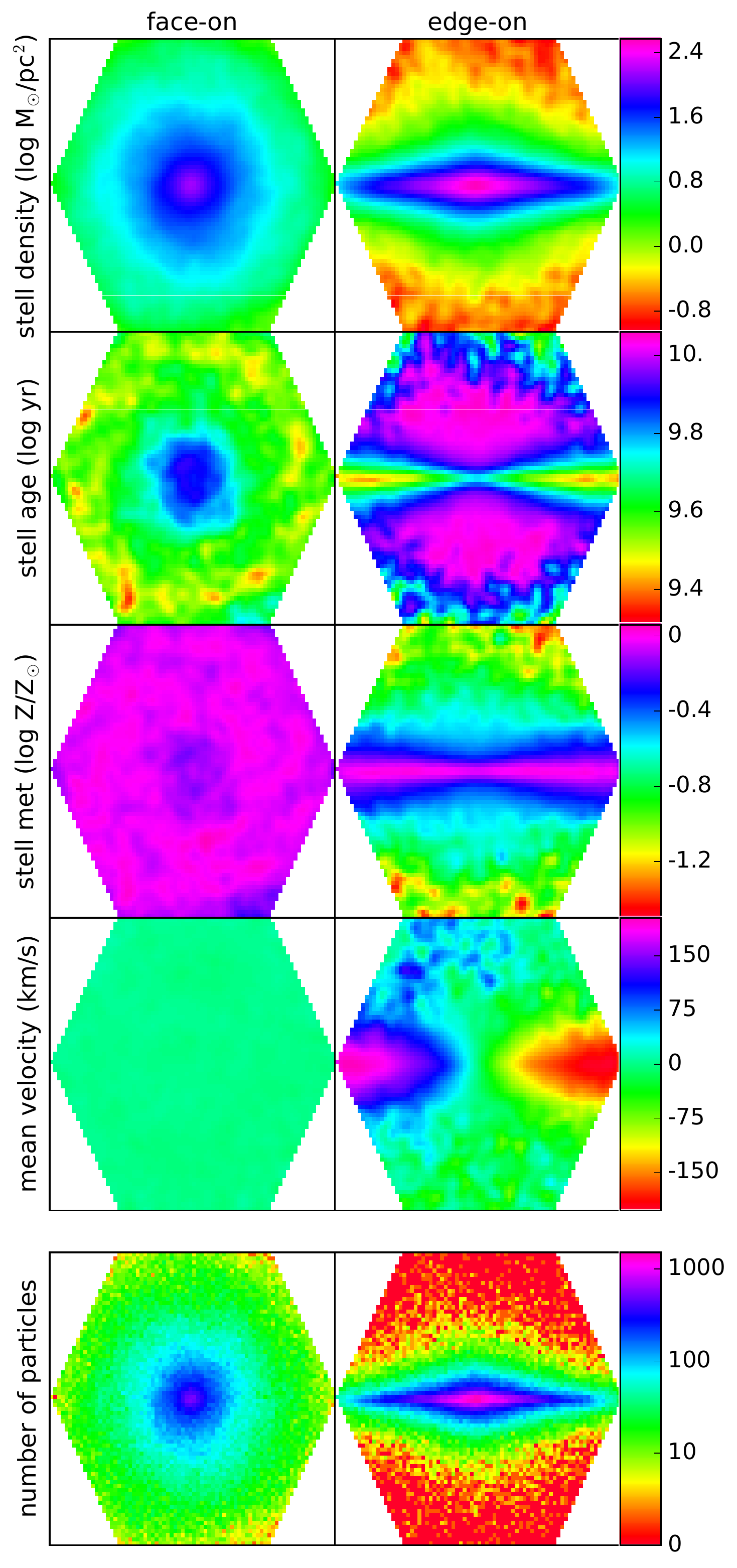}}
\caption{Spatially-resolved stellar properties of two simulated galaxies, 
D-MA\_0 (face-on) on the left and D-MA\_2 (edge-on) on the right. These maps 
show, from top to bottom, the stellar mass density, the mean 
luminosity-weighted ages and metallicities, the mean velocity along the 
line of sight, and the number of stellar particles
in each spaxel (in logarithmic colour scale).}
\label{fig_resolved_maps_stellar}
\end{figure}

%%%%%%%%%%%%%%%%%%%%%%%%%
%\end{comment}

In addition, we also provide spatially-resolved measurements of the
  quantities listed in Table~\ref{tab:list_stellar_properties}, considering
  the particles 
within each spaxel of our virtual CALIFA observations (see Section~\ref{sec:califa}) for three different orientations.

\begin{itemize}

\item { Stellar mass / surface density:} total stellar mass within
  the spaxel and stellar surface density
  
  %total mass within the spaxel, $\log \frac{M_*}{M_\odot}$, and surface density, $\log \frac{\Sigma_*}{\rm M_\odot/pc^2}$.

\item { Star formation rate:} calculated averaging the mass of stars
  formed in the last 100~Myr \citep{Kennicutt98}
  %in units $\text{M}_{\odot}~\text{yr}^{-1}$.

\item { Stellar age / metallicity:} 
mass- and luminosity-weighted means~\eqref{eq:age_mass_w} and~\eqref{eq:age_lum_w}

\item { Luminosity-weighted velocity \eqref{eq:v_mean} and
  velocity dispersion \eqref{eq:v_disp}.}
\item Star formation histories: stellar mass formed within 100 Myr bins
  ordered in lookback time.

\end{itemize}

Some examples of these \emph{product datacubes} (spatially resolved maps of intrinsic physical properties that are directly measured from the simulations) can be seen in Figure~\ref{fig_resolved_maps_stellar}.
They represent the final `solutions' to be recovered from the mock observational data.
{
Note that, for luminosity-weighted averages, we are adopting the usual definition of `intrinsic physical properties' that ignores the effects of radiative transfer.
}

%--------------------------------------------------------------------------
\section{Spectral energy distribution}
\label{sec:simulated_spectra}
%--------------------------------------------------------------------------

%%%%%%%%%%%%%%%%%%%%%%%%%%%%%%%%%
%\begin{comment}

\begin{figure*}
{\textbf{C-CS$^+$}\par\medskip\vspace{-0.1cm}
\includegraphics[height=.2\textheight]{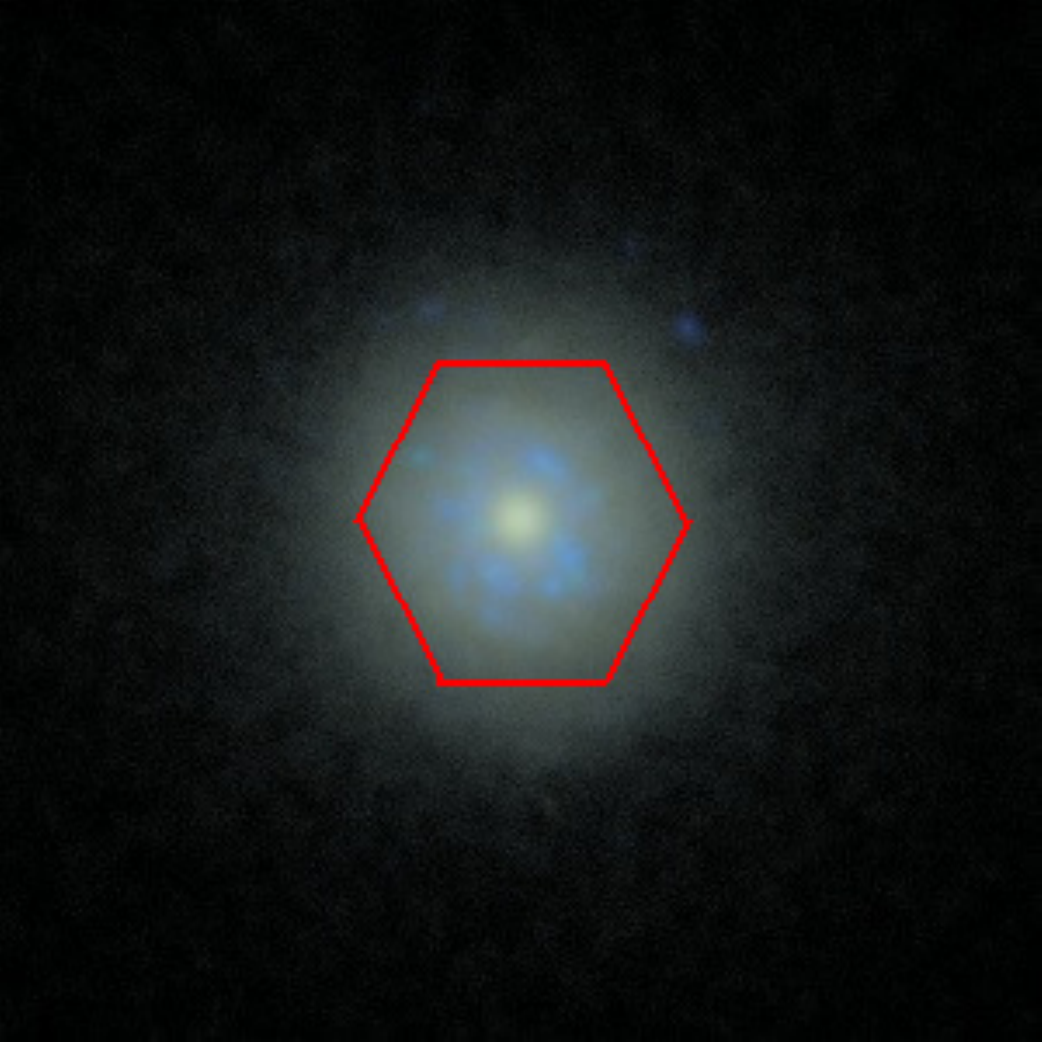}
\includegraphics[height=.2\textheight]{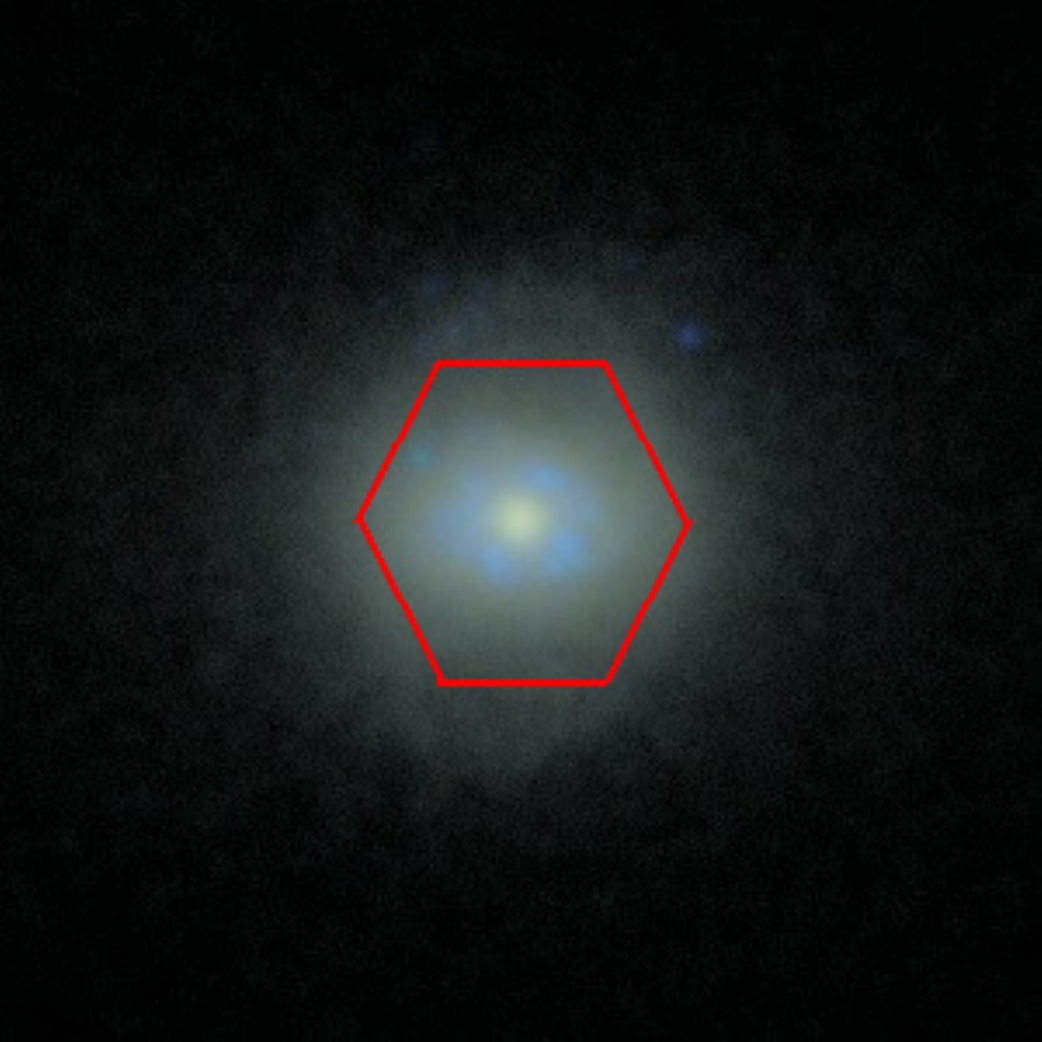}
\includegraphics[height=.2\textheight]{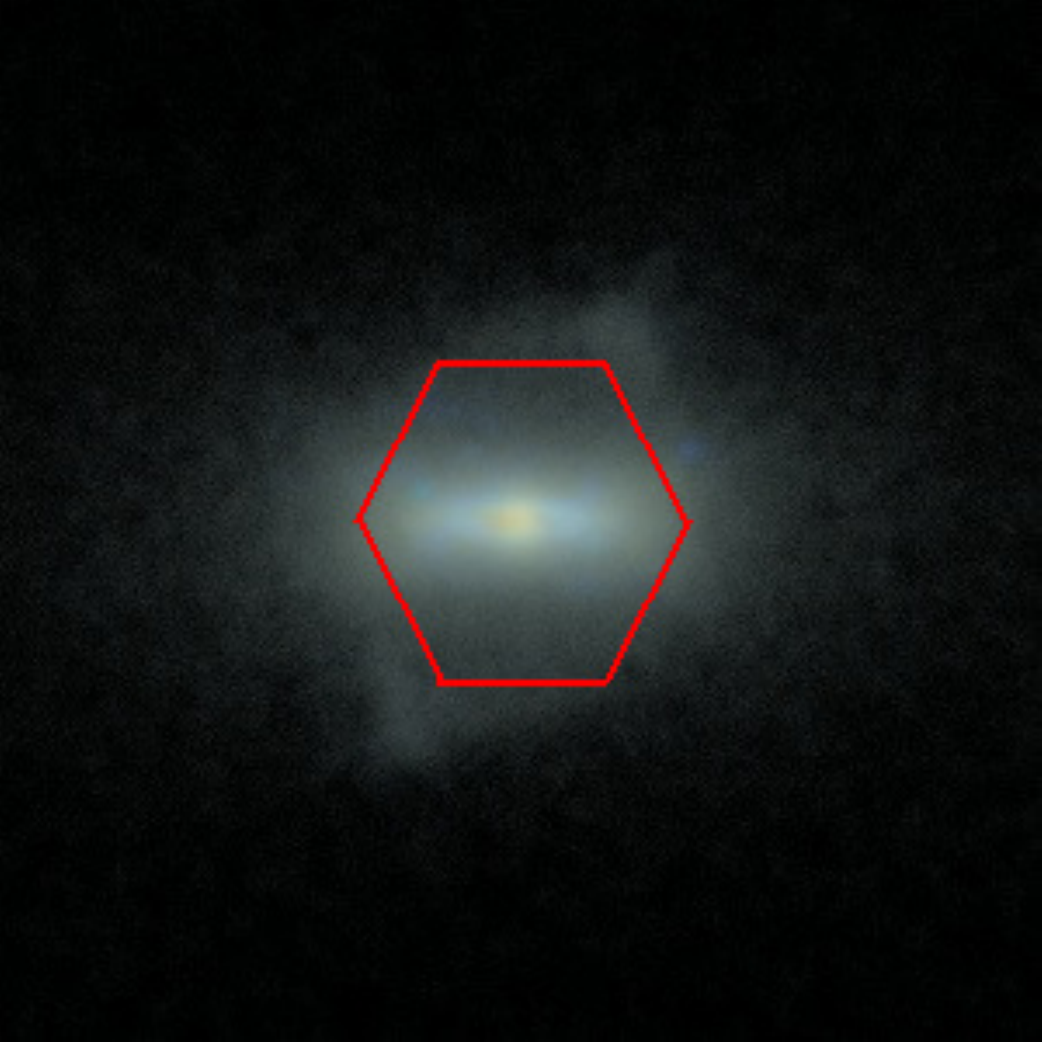}}
\vspace{0.2cm}

{\textbf{E-CS$^+$}\par\medskip\vspace{-0.1cm}
\includegraphics[height=.2\textheight]{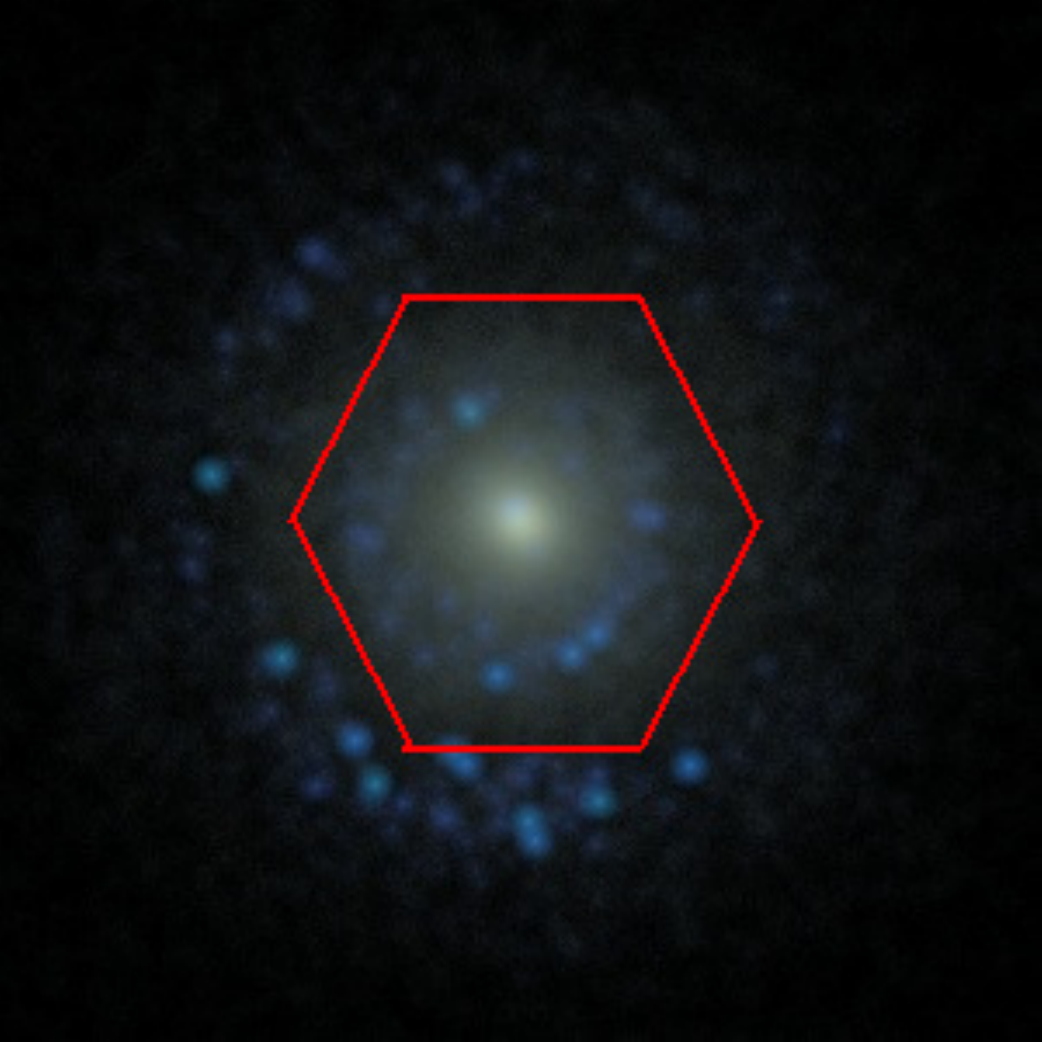}
\includegraphics[height=.2\textheight]{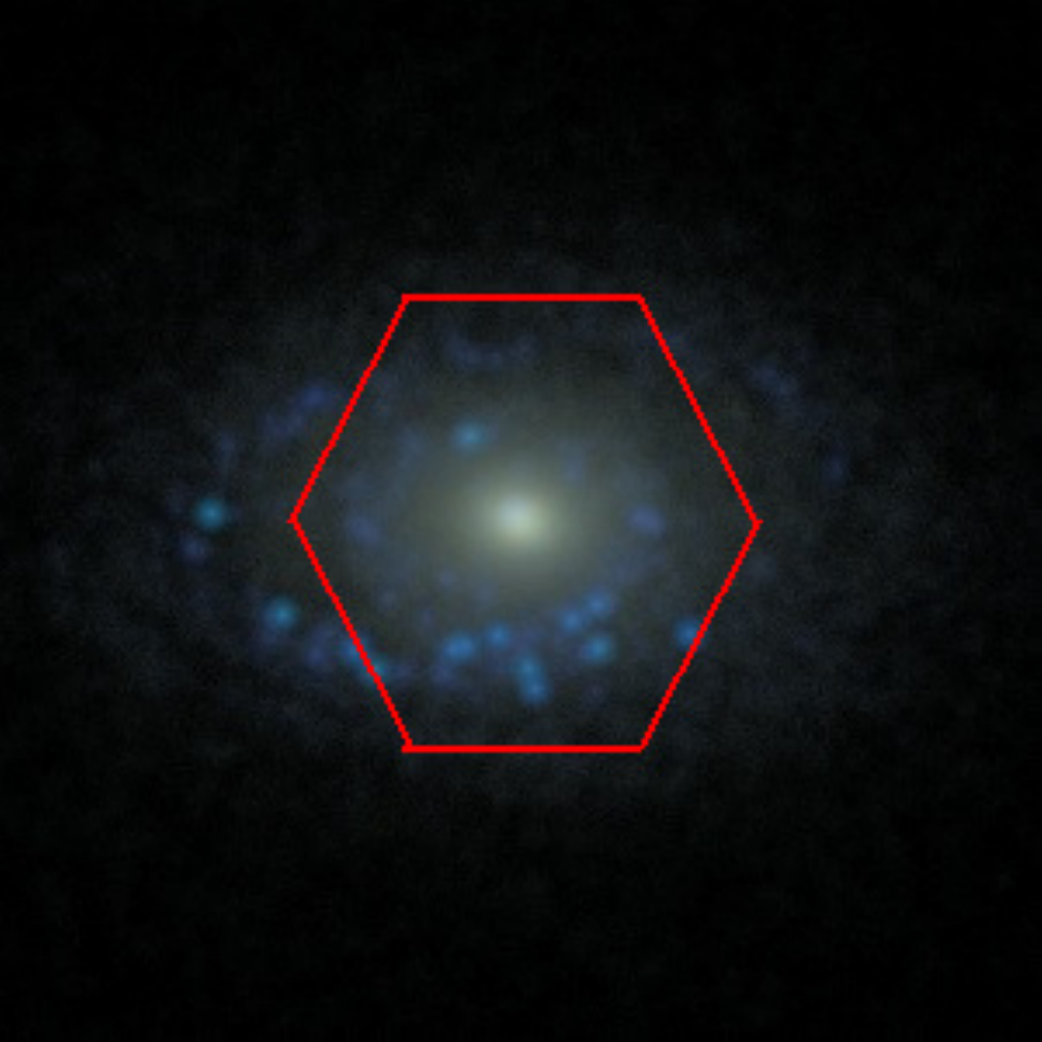}
\includegraphics[height=.2\textheight]{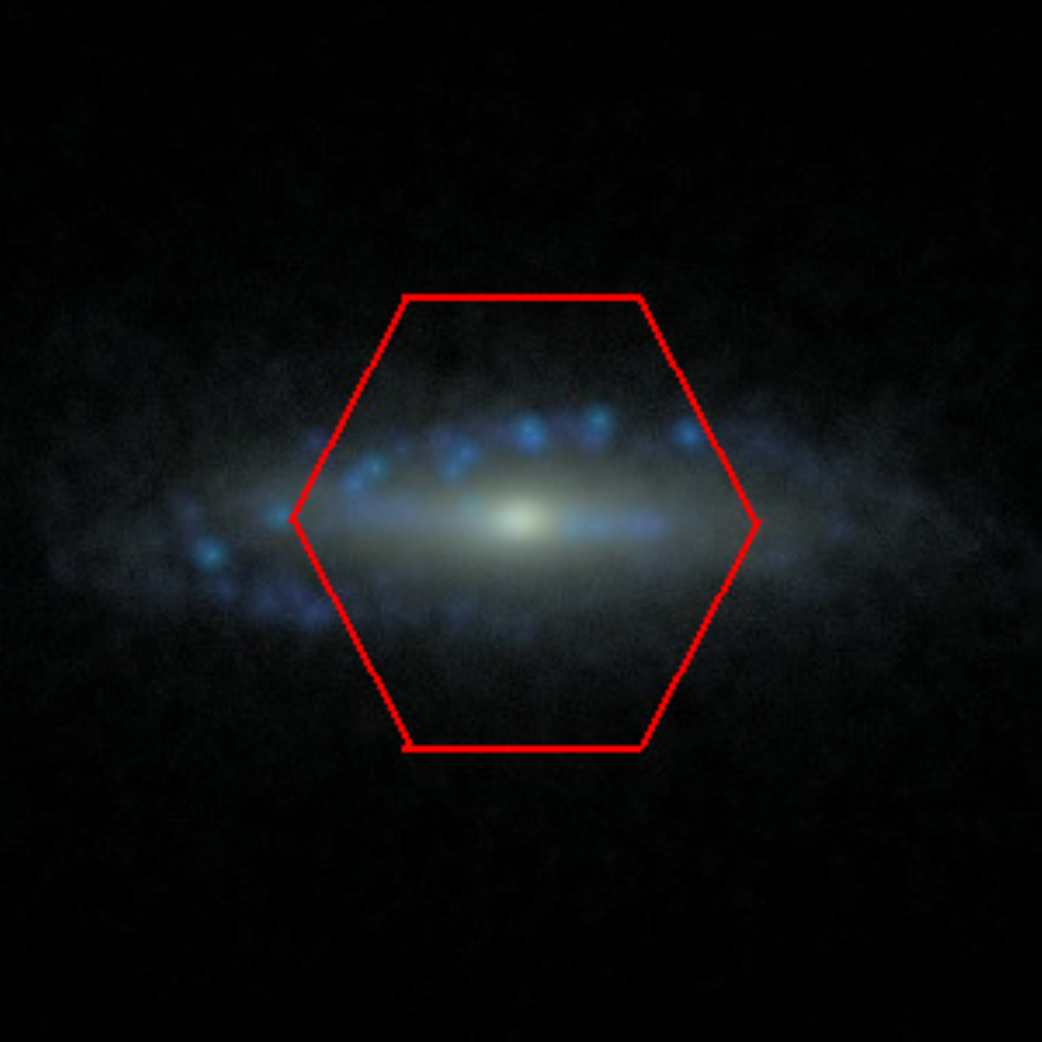}}
\vspace{0.2cm}

{\textbf{D-MA}\par\medskip\vspace{-0.1cm}
\includegraphics[height=.2\textheight]{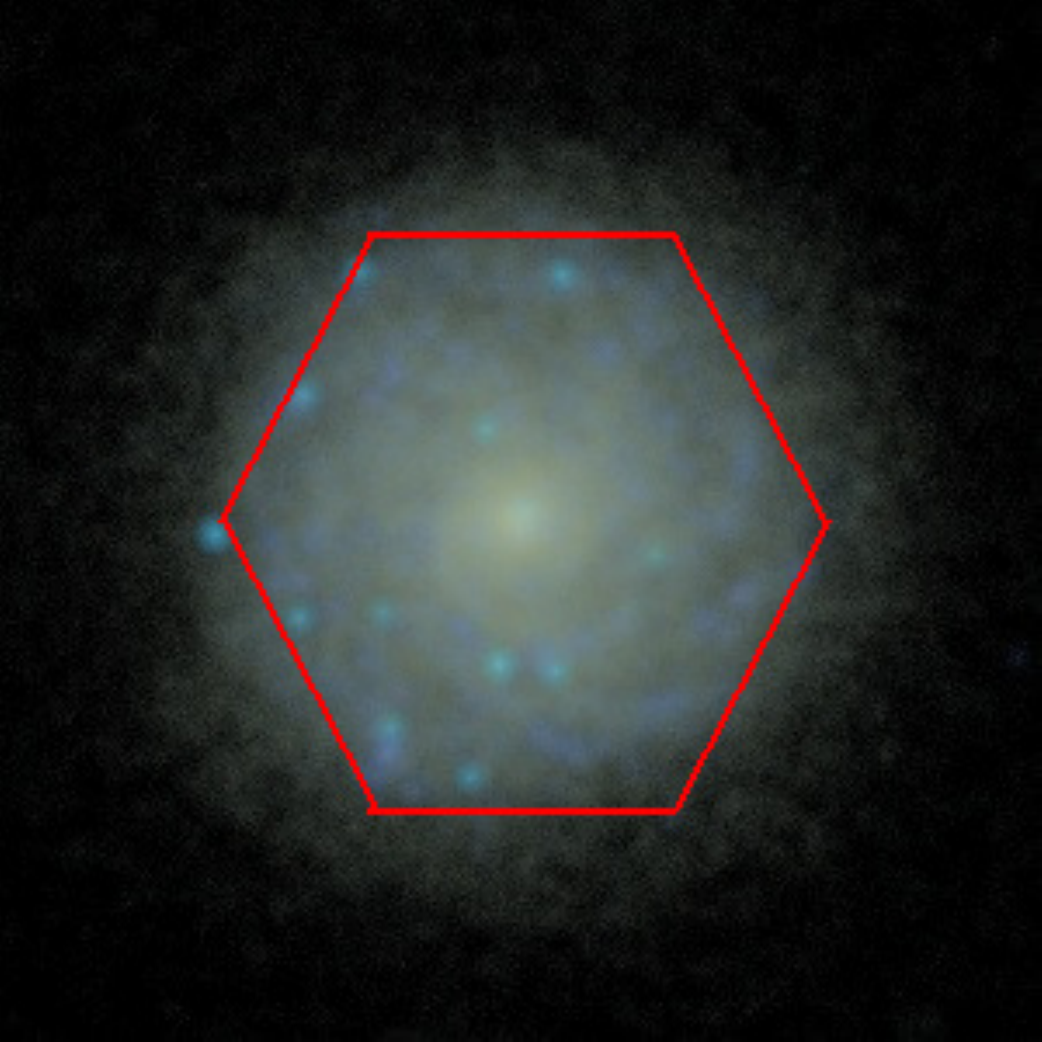}
\includegraphics[height=.2\textheight]{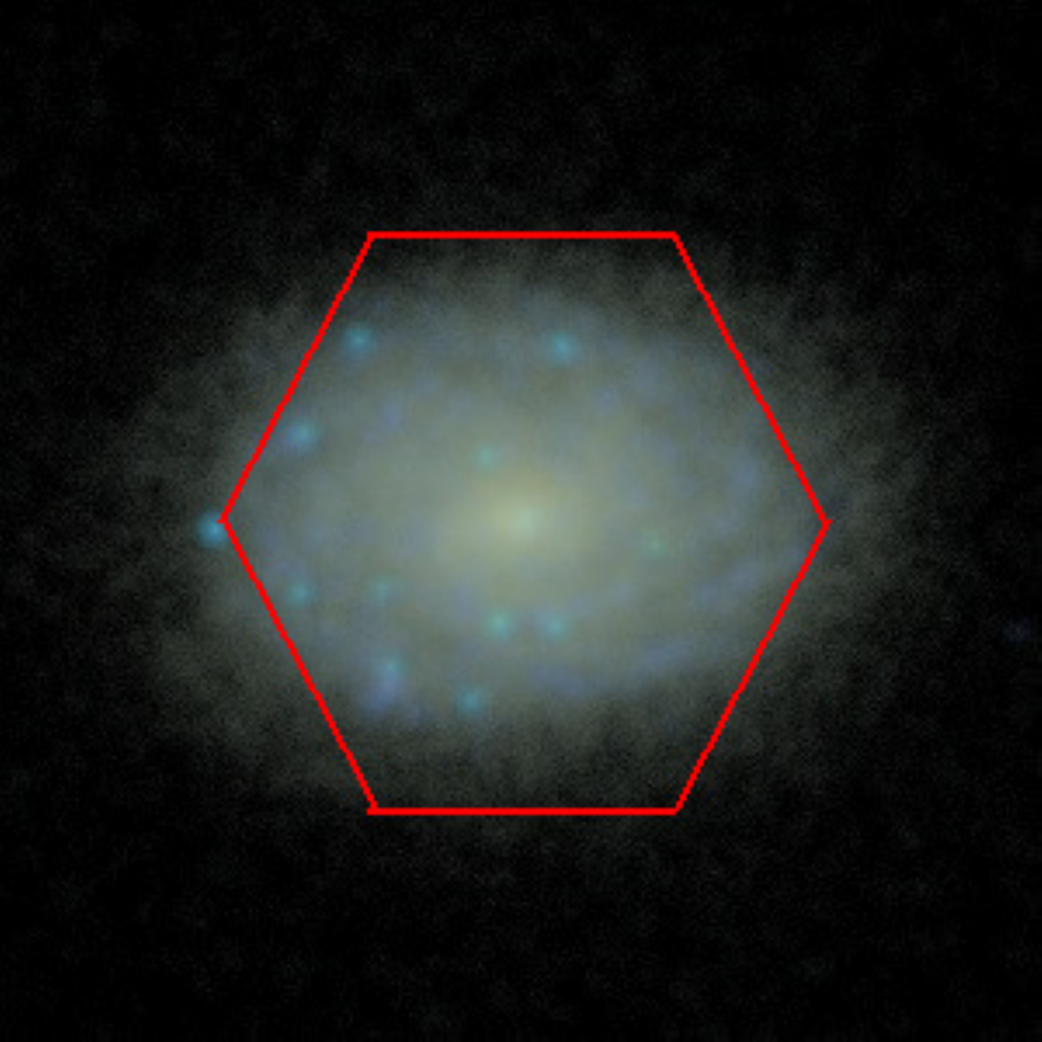}
\includegraphics[height=.2\textheight]{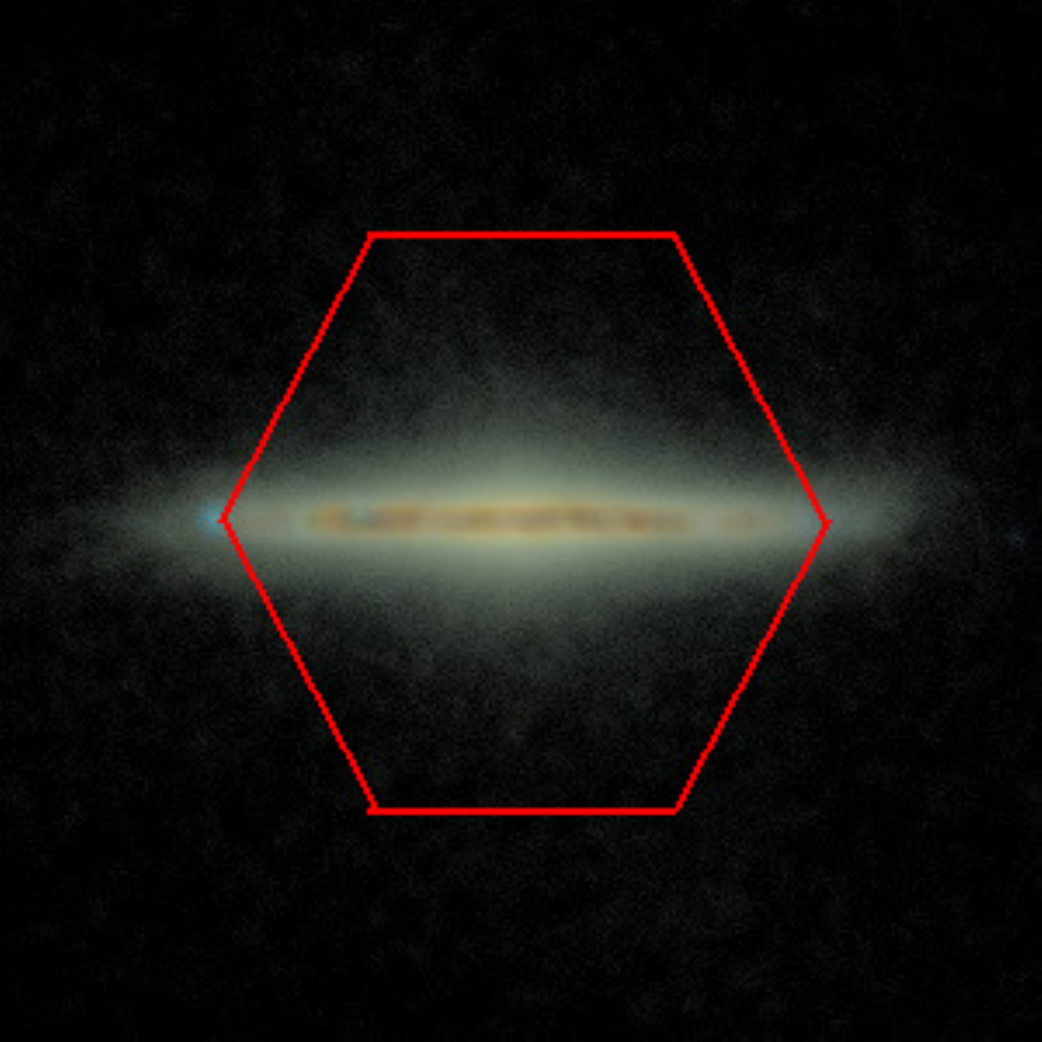}}
\vspace{0.2cm}
\caption
{Composite synthetic broadband images created in the $(u,r,z)$-bands 
using the \citet{Lupton04} composition algorithm for our three 
simulated galaxies (from top to bottom, C-CS$^+$, E-CS$^+$, and D-MA) in
a field of view of $60 \times 60$ kpc with $300 \times 300$ pixels. 
The orientations are, from left to right, face-on, $45^{\circ}$ and 
edge-on, labelled in the synthetic datacubes as \_0, \_1 and \_2
respectively. 
The red hexagon is the region of the simulations sampled by the CALIFA 
field of view, with physical sizes of $\sim 19, 27, 35$ kpc respectively 
for the 
C-CS$^+$, E-CS$^+$ and D-MA galaxies.}
 \label{fig:galaxy_images}
\end{figure*}

%\end{comment}

To generate the spatially-resolved spectral energy distribution of our 
simulated galaxies, we post-process the particles in the main halo in the
simulations' snapshots at redshift 
zero with the Monte Carlo radiative transfer code {\sc sunrise} 
\citep{Jonsson06, Jonsson10}.
{\sc sunrise} is able to self-consistently compute the 
emission and propagation of light in a dusty InterStellar Medium (ISM).
The resulting SEDs include the contribution
of stellar and nebular emission, dust absorption and scattering, and 
hence show stellar absorption features, emission lines, 
as well as the effects of kinematics. 
More details about the {\sc sunrise} input parameters used in this
work can be found in \citet{Guidi16}.

In the first step, the code assigns a specific spectrum to
each stellar particle depending on its age and metallicity, normalized to the mass of the particle.
For ages $> 10$ Myr,
spectra from the {\sc starburst99} Stellar Population Synthesis (SPS)
model (SB99, \citealt{Leitherer99}) are assigned.
They combine low-resolution stellar models ($\sim 20$-\AA\ sampling)
for wavelengths $\lambda < 3000$~\AA\ and $\lambda > 7000$~\AA, based on
the Pauldrach/Hillier stellar atmospheres\footnote{{This resolution is lower than the CALIFA sampling, and therefore the affected wavelength range has been labelled as bad pixels in the mock observations.}}, while the high-resolution
range ($3000-7000$~\AA, with $0.3$-\AA\ sampling) uses
the fully theoretical atmospheres by \citet{Martins05}.
On the other hand, young particles with age $\le 10$ Myr are
assigned a spectrum that takes into account the photo-dissociation 
and recombination of the gas, computed with the 1D photo-ionization code 
{\sc mappings III} \citep{Groves04,Groves08}.

In the radiative transfer stage, $\sim 10^7$ randomly-generated 
photon packets are propagated through the dusty ISM with a Monte Carlo 
approach (with a constant dust-to-metals ratio of 0.4, 
\citealt{Dwek98}) assuming a Milky Way-like dust curve with $R_V =3.1$
\citep{Cardelli89,Draine03}. 
In the calculation of the kinematic broadening of the lines only 
the radial velocity of the particles 
is taken into account, neglecting both the velocity dispersion intrinsic
in stellar clusters and the thermal broadening of the
gas emission lines.
This effect is certainly relevant for the intrinsic width of the emission
lines\footnote{For this reason, as well as for the scarcity of young
  stellar particles, we focus on the intensity of nebular emission lines
  and refrain from considering gas kinematics.}, and in principle it could
affect the detailed broadening of the stellar SED in those areas where
few particles are present in a given spaxel (see bottom panel of
Figure~\ref{fig_resolved_maps_stellar}).
Note that the {\sc sunrise} kinematic algorithm consistently 
takes into account the Doppler shift when the light is scattered 
by dust, which can also modify the line profiles
\citep{Solf91, Henney98, Grinin12}.

Finally, Monte Carlo photons are collected by} cameras 
placed around the simulated galaxies to obtain the 
SED in each pixel.
In our calculations we place cameras
with three different orientations (defined according to 
the alignment of the total angular 
momentum of the stars with the $z$ direction) for each galaxy, 
respectively face-on, $45^{\circ}$ and edge-on (see
Fig.~\ref{fig:galaxy_images} for the \{$u,r,z$\}-band colour-composite 
images).
In order to reduce the random noise introduced by the {\sc sunrise} 
Monte Carlo algorithm, the radiative transfer process described above 
is run ten times for 
each galaxy, changing 
only the random seeds, and the resulting spectra are averaged over the 
ten different random realizations. 
In this way we are able to reach a 
`signal-to-noise' S/N (where N is the standard deviation over the ten
realizations) of $\sim 300-400$ in the central spaxels and 
S/N$\sim 8 - 10$ in the outer regions, which is negligible compared to 
the typical values of the S/N in the CALIFA spectra that we aim to mimic
\citep{Sanchez12}.

%%%%%%%%%%%%%%%%%%%%%%%%%%%%%%%%%%
%\begin{comment}

\begin{figure*}
\centerline{\includegraphics[width=.8\textwidth]{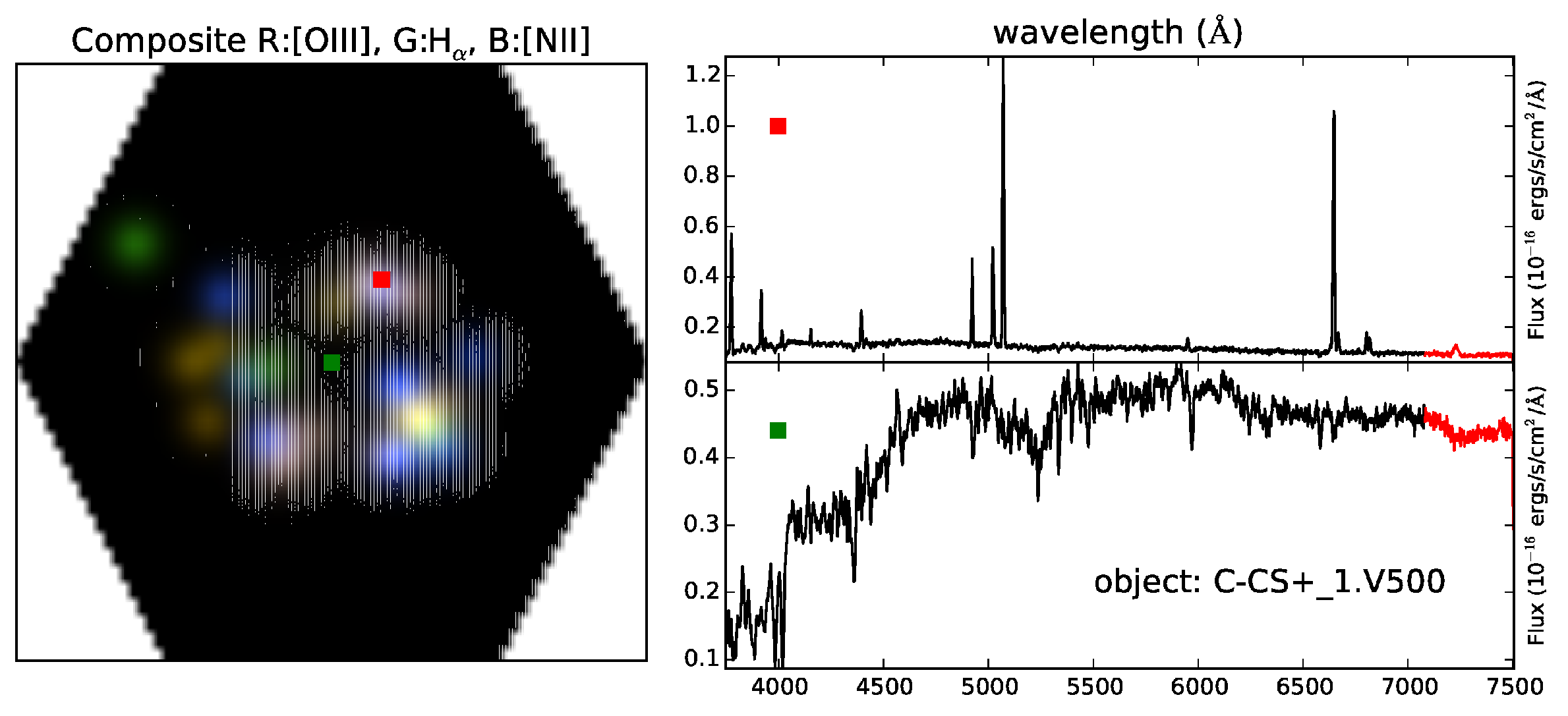}}
\caption{Left panel: RGB image of the [OIII]5007, H$\alpha$ and [NII]6584 \
emission lines for the galaxy C-CS$^+$\_1. Right panels: synthetic spectra 
in the spaxels corresponding to the red and green squares in the RGB image. 
In the upper right panel, the spaxel samples a nebular region (red square), while the 
lower right panel shows a V500 spectrum containing
only stellar emission (green square). The part of the spectrum generated with the 
low-resolution stellar model (Sec.~\ref{sec:simulated_spectra}) is marked in 
red in the plot. 
}
\label{fig:RGB_plot}
\end{figure*}

\begin{figure}
\centerline{\includegraphics[width=.4\textwidth]{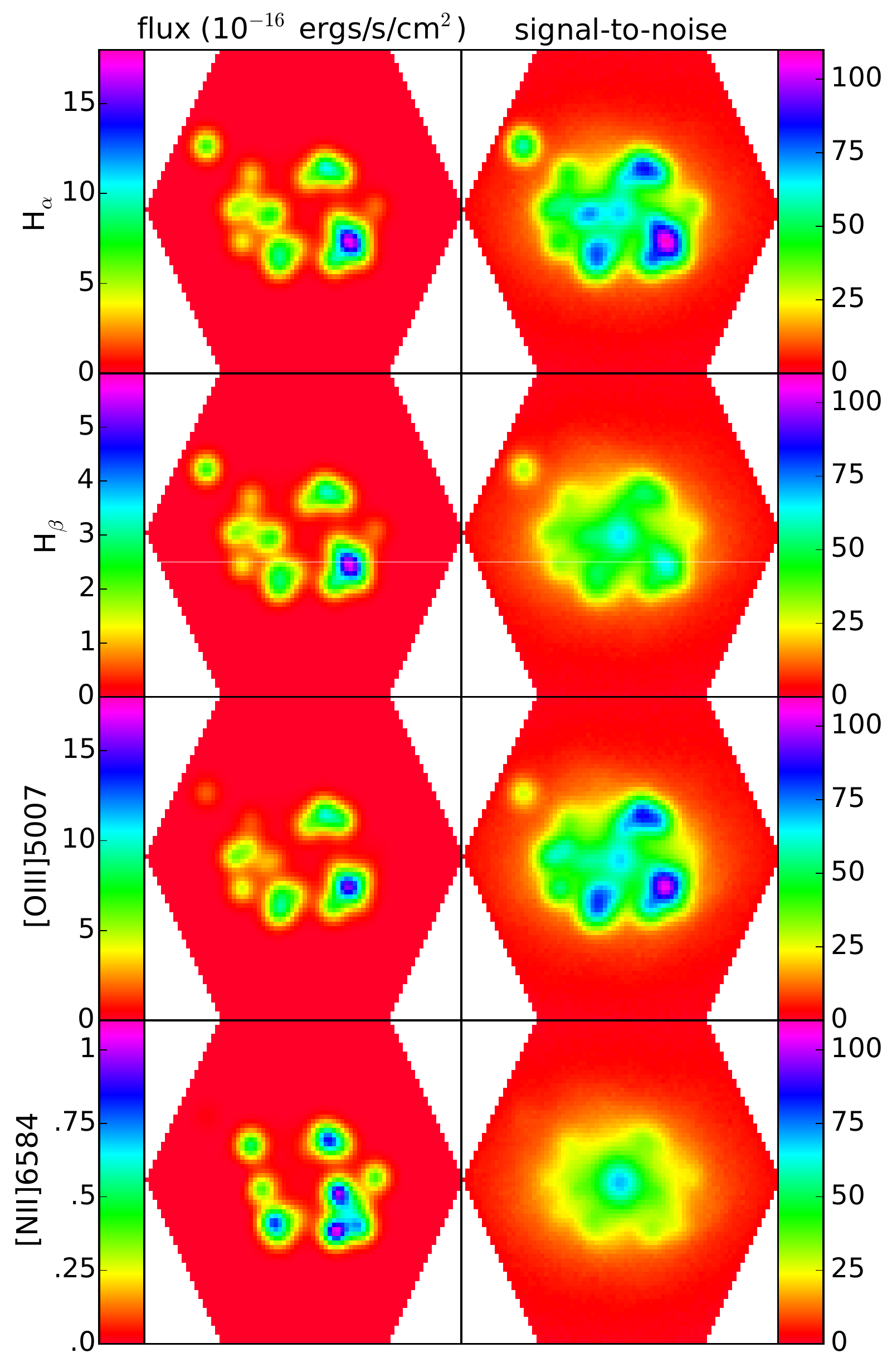}}
\caption{Line intensity (left) and Signal-to-Noise (right) maps of the four BPT
lines \citep{Baldwin81} for the galaxy C-CS$^+$\_1. The S/N in every spaxel 
is obtained as the ratio between the mean signal and noise in the 
wavelength range of 
the corresponding emission line given in Table~\ref{tab:list_emission_lines}.}
\label{fig:resolved_maps_nebular}
\end{figure}

%%%%%%%%%%%%%%%%%%%%%%%%%%%%%%%%%%%%%%%%%
%\end{comment}

As part of the product datacubes, we provide the predicted SEDs of our galaxies, free from instrumental effects, as well as resolved maps of some spectral features derived from these theoretical datacubes.
To obtain measurements of the properties of the 
stellar and nebular spectra without using any observational 
algorithm to separate the two spectral components (which may introduce 
many caveats and uncertainties),   
we additionally generate {\it stellar-only} synthetic datacubes 
following the same procedure described above,
replacing the {\sc mappings} spectra with {\sc starburst99} templates.

We generate the following maps:

\begin{itemize}

\item { Lick indices:}
the strength of the Lick indices (Table~\ref{tab:stellar_absorption_indices} in Appendix~\ref{sec:product_datacubes}) is derived from the 
stellar-only datacubes, which have $2$-\AA\ sampling.

\item{ Nebular emission line intensities:}
{ the fluxes of different emission lines 
(Table~\ref{tab:list_emission_lines}) are measured 
after subtracting the stellar-only datacubes to the synthetic ones in order
to take into account stellar absorption features in the calculation.
Since the dust around young stars in the stellar-only datacubes is 
neglected, while is present at the sub-resolution level with {\sc mappings},
for each line we renormalize the stellar-only spectra with the continuum
level of the full cube at the lower bound of each line, and we compute 
the total flux between the lower and upper bounds of the lines
after subtracting the continuum (the offsets 
between full and stellar-only cubes are indeed quite 
small, of the order $\sim  0.5 - 2\%$ of the flux).
It is important to emphasize here
that the nebular emission in the datacubes is limited to  
the stellar particles younger than 10 Myr (HII regions), and we do 
not count on any other sources of ionizing photons\footnote{{Actually, nebular emission lines are only produced at the location of young stellar sources. UV photon leakage is not considered, and the diffuse ISM merely absorbs and scatters the nebular emission. At variance with real galaxies, it does not produce any light by itself.}}.
The units are $10^{-16}$~erg~s$^{-1}$~cm$^{-2}$~\AA$^{-1}~\text{spaxel}^{-1}$.
}

\end{itemize}

It should be noted that these quantities are measured from simulated spectra including the effects of dust and kinematics.
Lick indices are defined over a short wavelength range, and therefore they are not expected to be strongly affected by dust extinction, although differential effects (highly dependent on the numerical resolution and assumed ISM physics of the simulations) may certainly play a role for realistic star formation histories \citep{MacArthur05}.
They also depend on
the velocity dispersion at which they 
are measured (see e.g. \citealt{Sanchez-Blazquez06, Oliva-Altamirano15}). 
In IFS observational studies the spectra in each spaxel are usually 
broadened to a single velocity dispersion prior to the measurement 
of the Lick indices, in order to consistently compare them with models 
with the same dispersion \citep[e.g.][]{Wild14}.
In our product datacubes we do not change the broadening of the 
absorption lines, 
since this procedure introduces additional uncertainties in 
the analysis. The spaxel-by-spaxel velocity dispersion is provided in the 
GALNAME.stellar.fits files (Sec.~\ref{sec:product_datacubes}) and can 
be used to tune the fitted models.

Our datacubes provide the nebular line intensities
  considering both the extinction within the nebula (implicit in the
  MAPPINGS templates) as well as absorption and scattering in
  the ISM (computed by {\sc sunrise}).
In Figure~\ref{fig:RGB_plot} we show an RGB image of the intensities (derived 
from the nebular maps) of the [OIII]5007, H$_{\alpha}$ and [NII]6584
emission lines, together with spectra in two different 
spaxels in the synthetic datacubes, one containing nebular emission and the other 
only stellar light.
An example of these maps is given in Figure~\ref{fig:resolved_maps_nebular}, 
where we show for one of our simulated galaxies the intensities of the BPT 
 \citep{Baldwin81} emission lines  
(H$_{\alpha}$, H$_{\beta}$, [OIII]5007, [NII]6584), and 
the corresponding signal-to-noise maps (see Section~\ref{sec:califa}).

  At variance with the intrinsic galaxy properties discussed in
  Section~\ref{sec:simulations}, we consider that Lick indices and nebular
  emission line intensities are properties of the SEDs, including the effects
  of geometry, kinematics, and radiative transport.
  These quantities (which may in principle be quite different from the sum
  of the intrinsic emissivities) are to be recovered from imperfect data,
  affected by the observational effects described below.

%--------------------------------------------------------------------------
\section{Mock observations}
\label{sec:califa}
%--------------------------------------------------------------------------

In this section we describe how we convert the output of 
the {\sc sunrise} radiative transfer algorithm into synthetic IFS observations 
mimicking the CALIFA survey \citep{Sanchez12, Garcia_Benito15}.
We briefly summarise the technical properties of the survey and explain
  how the main steps of the observation procedure are reproduced.
A detailed description of the products that we make publicly available is provided in Appendix~\ref{sec:selgifs}.

\begin{table}
\begin{center}
\begin{tabular}{cccc}
\hline
Object & Redshift & Luminosity distance & Physical size  \\
 &  & [Mpc] & of the spaxels [kpc]  \\
\hline
C-CS$^+$ & 0.013 & 58.1 & 0.25 \\
E-CS$^+$ & 0.018 & 80.8 & 0.35 \\
D-MA  & 0.024 & 108.2 & 0.45 \\
\hline
\end{tabular}
\end{center}
\caption{Redshift, luminosity distance and physical size covered by the 
spaxels in our synthetic CALIFA observations.}
\label{tab:redshift}
\end{table}

\begin{table*}
\begin{center}
%\scalebox{0.95}{
\begin{tabular}{lcccccccc}
\hline
Setup & $N_{\alpha}$ & $N_{\delta}$ & $N_{\lambda}$ & $\lambda$ (\AA) & $d_{\lambda}$ (\AA) & $\delta_{\lambda}$ (\AA) & $\sigma_0$ & $I_0$ \\
\hline
V500   & 78 & 73 & 1877 & $3749-7501$ & 2.0 & 6.0 & 0.29 & 20.8 \\
V1200  & 78 & 73 & 1701 & $3650-4840$ & 0.7 & 2.3 & 0.64 & 25.9 \\
\hline
\end{tabular}
%}
\end{center}
\caption
{Sizes of the simulated datacubes in the spatial and spectral dimensions 
($N_{\alpha}, N_{\delta}, N_{\lambda}$),
  { wavelength range}, spectral sampling, and spectral resolution ($\lambda$, $d_{\lambda}, \delta_{\lambda}$) { of each instrumental setup}, and
  best-fitting parameters ($\sigma_{0}, I_{0}$) in equation~\eqref{eq:noise}
{ for the row-stacked spectra (RSS) of individual pointings in the CALIFA DR3.}}
\label{tab:data_format}
\end{table*}

CALIFA observations were taken with the
Potsdam Multi-Aperture Spectrograph \citep[PMAS,][]{Roth05},
mounted on the Calar Alto 3.5-m telescope, 
utilizing the large hexagonal field of view offered by the
PPak fibre bundle \citep{Verheijen+04, Kelz+06}.
The final CALIFA Public Data Release (DR3, \citealt{Sanchez16_1})\footnote{\url{http://califa.caha.es/DR3}}
consists of 667 galaxies. 
Sample selection \citep{Walcher+14} and observing strategy
  \citep{Sanchez12} aimed for
a filling factor of 100\% up to $\sim 2.5$ effective radii $R_{e}$.
Datacubes have a field of view of $[76-78]''$ in right ascension and $[71-73]''$ in declination
(depending on the observing conditions and the precise disposition of the dithering pattern for each object)
sampled by $1''$ spaxels ($\sim 1$-kpc physical size).
To reproduce these characteristics, we adopt a fixed $78''\times 73''$ configuration,
we derive the half-light radius ($R_{50}$) in the $r$-band for every simulated galaxy,
and then we adjust their redshift/distances (see Table~\ref{tab:redshift})
so that $78''$ correspond to $4 \, {R_{50}}$.

The wavelength range, sampling, and resolution of the two spectral setups of
the CALIFA survey \citep[V500 and V1200;][]{Sanchez12, Garcia_Benito15} are
quoted in Table~\ref{tab:data_format}.
% The low-resolution V500 mode covers the red range ($3749 - 7501$~\AA) with sampling 
% $d_{\lambda 500}=2.0$~\AA\ and full-width half-maximum (FWHM) resolution $\delta_{\lambda 500}=6.0$~\AA,
% whereas the blue ($3650 - 4840$~\AA) mid-resolution setup V1200 features
% $d_{\lambda 1200}=0.7$~\AA\ and $\delta_{\lambda 1200}=2.3$~\AA\ .
Due to the implementation of stellar atmospheres in {\sc sunrise}, (Section~\ref{sec:simulated_spectra}),
we generate two radiative transfer simulations (one with high resolution and kinematics for $\lambda=3000 - 7000$~\AA\ and another with low-resolution and no kinematics for $\lambda = 7000-7600$~\AA) for each galaxy. Then, we concatenate
the SEDs at $\lambda =  7000$~\AA,
apply the redshift of the object, and resample the spectra to the
appropriate $d_\lambda$ for each setup.
The regions of the V500 spectra generated with the low-resolution stellar model are flagged 
as bad pixels (see Section~\ref{sec:datacubes}), and they should not be used for SED fitting 
analysis\footnote{Notice that when we redshift our synthetic spectra 
we reduce the range of bad pixels, starting from the wavelength
$\sim 7090 - 7170$~\AA\ 
depending on the redshift of the object.}.

These datacubes are convolved with a Gaussian kernel of $3.25''$ FWHM
that adds in quadrature
the response of the $3''$-diameter fibres of the PMAS/PPak instrument
and an atmospheric seeing of $1.25''$, typical of a standard night at CALAR
ALTO observatory.
Then, they are `observed' three times, positioning the 331 fibres of the
instrument according to the dithering pattern followed in the CALIFA
survey \citep{Sanchez07, Sanchez12},
and the resulting $3 \times 331 = 993$ row-stacked spectra (RSS) are
convolved with another Gaussian that accounts for the spectral resolution
$\delta_\lambda$ (Table~\ref{tab:data_format}) of each grating.

%%%%%%%%%%%%%%%%%%%
%\begin{comment}

\begin{figure}
\includegraphics[width=.48\textwidth]{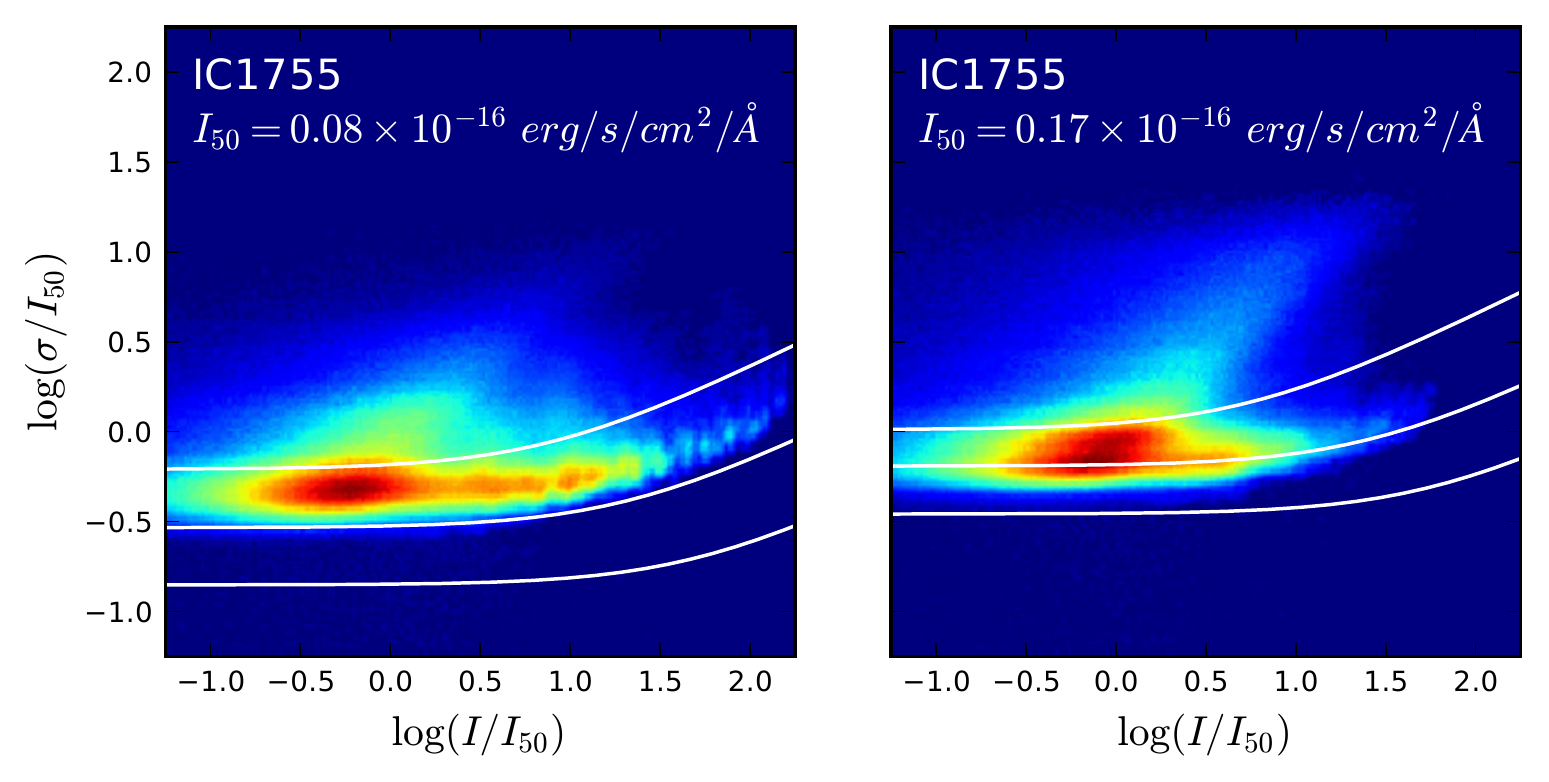}
\includegraphics[width=.48\textwidth]{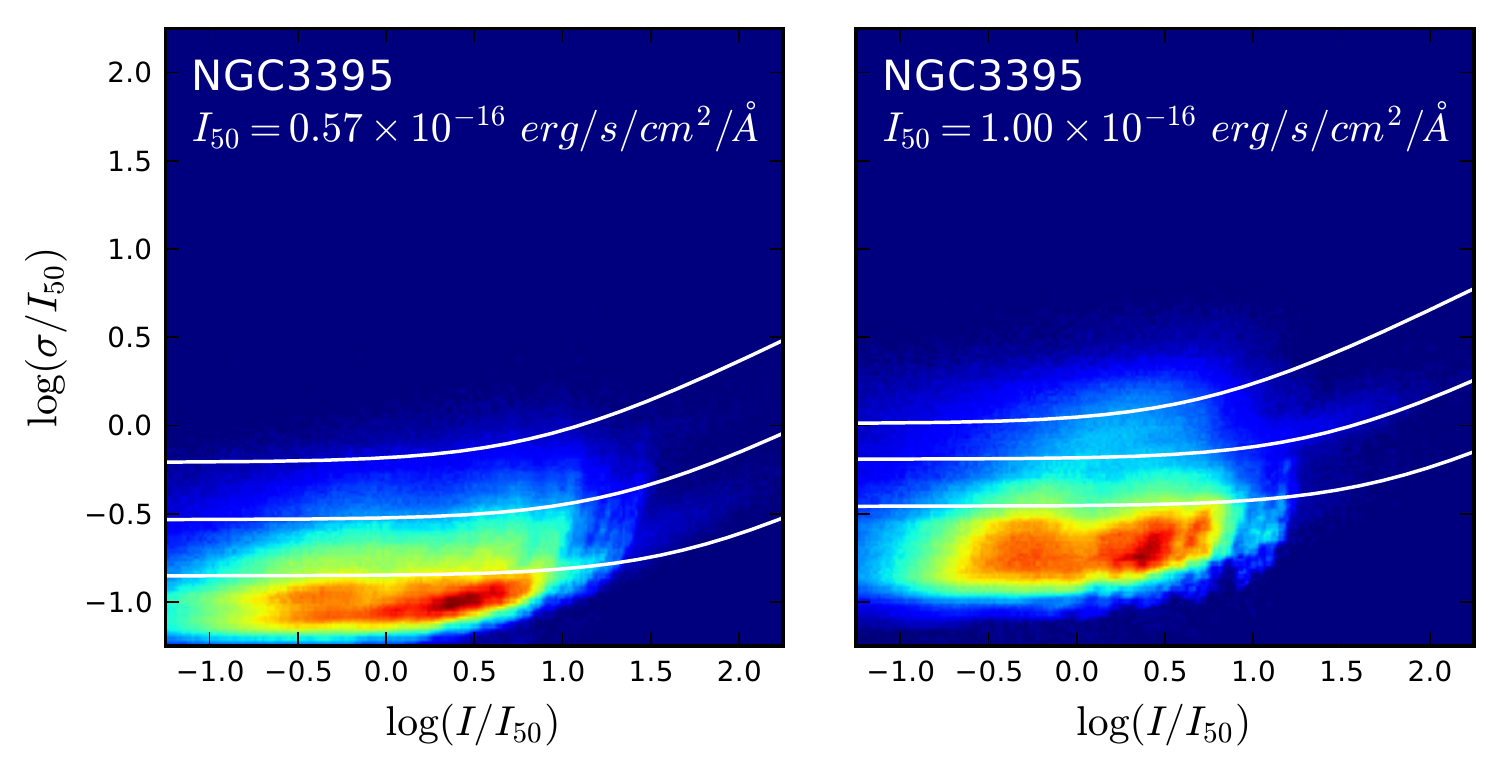}
\caption
{ RSS noise characterization. Colour maps showing the distribution of the noise $\sigma$ relative to the 
signal $I$, normalized to the median value of the signal $I_{50}$ 
(in logarithmic scale) for two RSS datacubes. 
Left (right) panel displays the values 
corresponding to the V500 (V1200) setup. 
Middle white solid line shows the
best-fit (median) curve for the $367\times 3$ CALIFA RSS.
The fitting parameters ($\sigma_{0}$ and 
$I_{0}$) are given in Table~\ref{tab:data_format}.
Solid white lines above and below represent the 90 and 10 percentiles of the parameter distribution.
Colour scale corresponds to the number of pixels 
($\sim 10^7$ in for every configuration).} 
\label{fig_noise_RSS}
\end{figure}

\begin{figure}
\includegraphics[width=.48\textwidth]{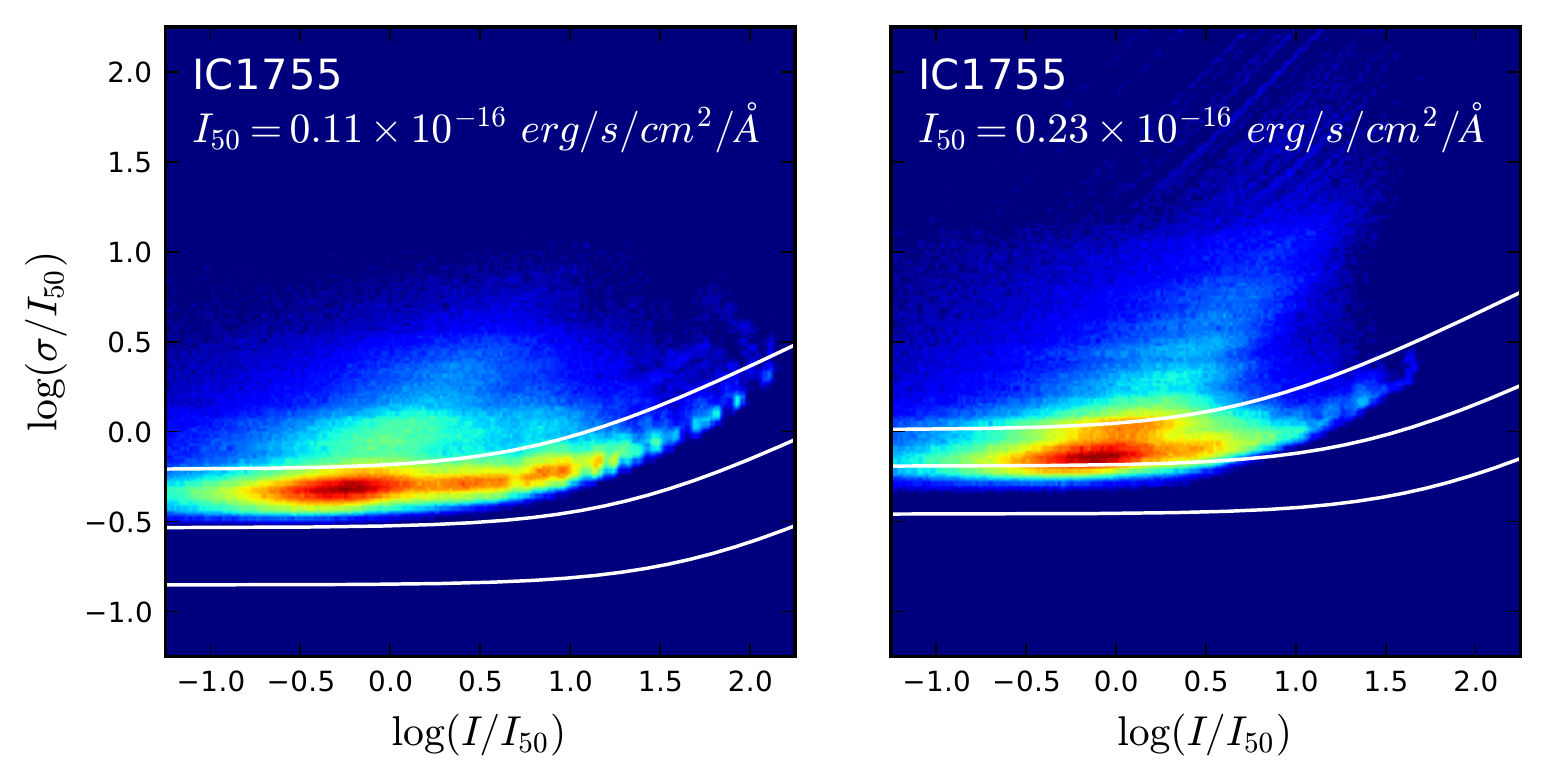}
\includegraphics[width=.48\textwidth]{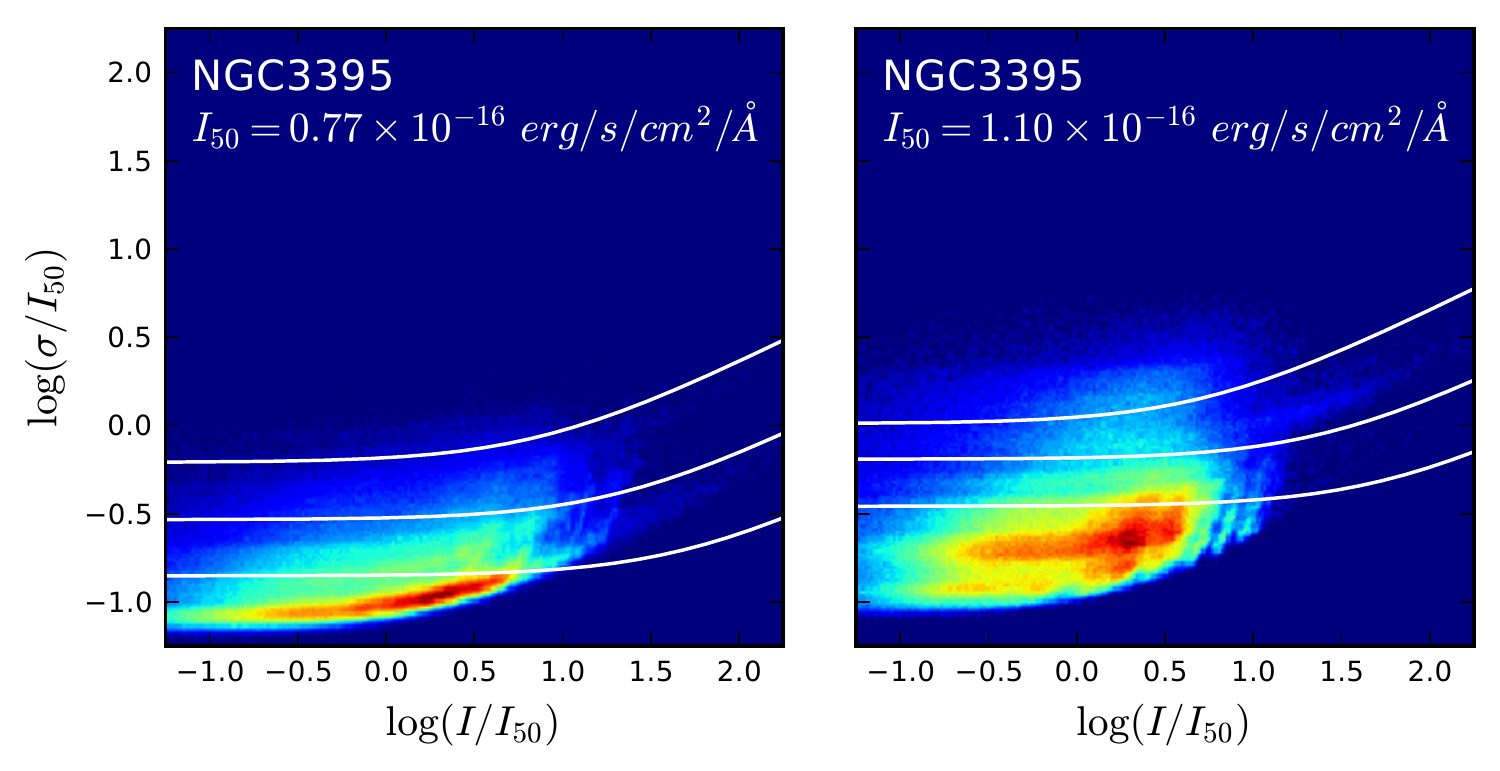}
\includegraphics[width=.48\textwidth]{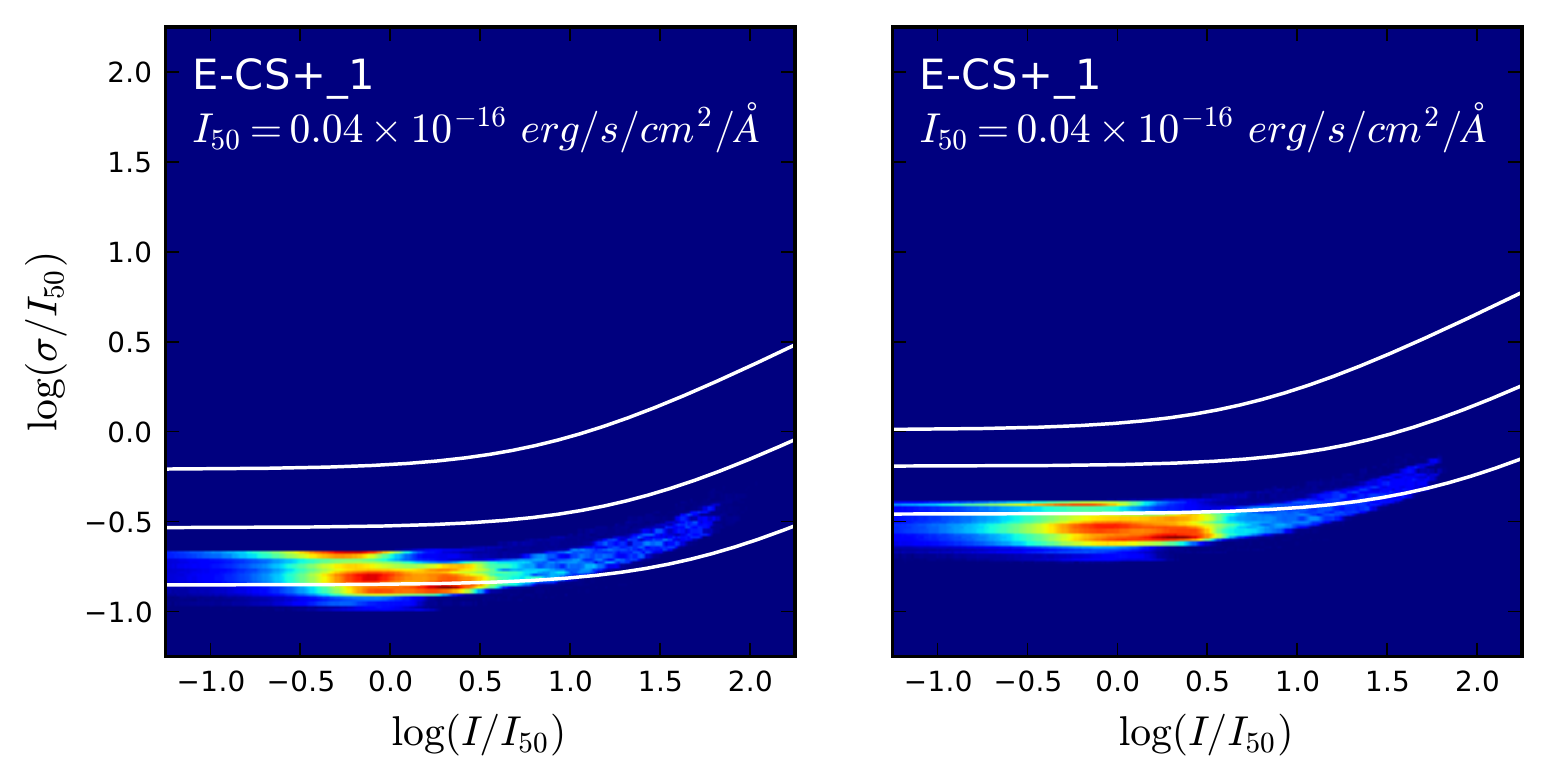}
\caption
    { Datacube noise. The plot is identical to that of
      Figure~\ref{fig_noise_RSS}. The upper two panels represent
      the noise distribution of the final datacubes corresponding to the
      two objects depicted in the previous figure. The lower panel
      corresponds to E-CS$^+$\_1, one of our synthetic datacubes. Our fit
      to the RSS data, prior to dithering and interpolation (same white
      lines as in Figure~\ref{fig_noise_RSS}) is shown to guide the eye.}
\label{fig_noise_datacubes}
\end{figure}

%%%%%%%%%%%%%%%%%
%\end{comment}

In order to account for detector noise, we selected galaxies from DR3 that were observed using both setups (V500 and V1200) and for which the quality flags indicate that the released cubes have only minor problems (see Section~6.4 of DR3). This sample consists of 367 galaxies out of the 667 forming the CALIFA DR3 \citep{Sanchez16_1}. We have analysed the raw-stacked spectra (RSS) corresponding to the three pointings of each object.
The results show that 
the dependence of the noise with the intensity of the signal, characteristic 
of charge-coupled devices, can be modelled with a simple parametric formula
\begin{equation}
\label{eq:noise}
\sigma_{\rm n} = \sigma_0\sqrt{1+\frac{I_{\rm n}}{I_0}}
\end{equation}
where $I_{\rm n}$ and $\sigma_{\rm n}$ refer to the observed intensity $I$ and errors $\sigma$ provided in the 
CALIFA datacubes normalized to the median value of the intensity, 
$\sigma_{\rm n}= \sigma /I_{50}$ and 
$I_{\rm n} = I / I_{50}$. 
The use of normalized errors and fluxes allows for a uniform 
object-independent unit-free characterization of the noise.
Our analysis also reveals that the detector noise does not depend on 
wavelength, except for the expected edge effects.

We fit the free parameters of
  equation~\eqref{eq:noise} to the data and then we take the median
value of the results over the 367 objects considered.
Best-fitting values are summarized in Table~\ref{tab:data_format}.
The panels of Figure~\ref{fig_noise_RSS} show the relation between $\sigma_{\rm n}$ and $I_{\rm n}$ for two RSS datacubes selected from the CALIFA sample. These objects are chosen such as their fitted $\sigma_{0}$ and $I_{0}$ values lie below and above the 10 and 90 percentiles in the sample respectively (i.e. extreme cases). White solid lines represent, from top to bottom, the 90, 50 (median), and 10 percentiles. 
These average relations inferred from CALIFA RSS data are used to add random Gaussian noise and set the errors of our synthetic RSS files (three pointings per simulated galaxy and orientation) containing 331 spectra each.

In a final step, each set of three RSS files is combined into a single interpolated datacube using version V1.5 of the CALIFA pipeline \citep{Garcia_Benito15}, following exactly the same procedure as in real observations.
Thus, our mock datacubes fully account for the effects of the dithering scheme,
yielding a final PSF of $\sim 2.5''$ FWHM \citep{Sanchez12} and introducing strong correlations between the noise of adjacent spaxels, as
shown in \citet{Husemann+13, Garcia_Benito15}
for real data.
An example of the errors assigned by the CALIFA pipeline to the final datacubes and to one of our synthetic ones are plotted in Figure~\ref{fig_noise_datacubes}.

%%%%%%%%%%%%%%%%%%%
%\begin{comment}

%--------------------------------------------------------------------------
\section{Science cases}
\label{sec:science}
%--------------------------------------------------------------------------

Although the main aim of this manuscript is to describe the SELGIFS Data Challenge, and provide all the necessary details about the data set so that it can be meaningfully explored by the community, let us briefly illustrate here the kind of scientific questions that could be address by means of a toy example: the ability of different methods to recover the main physical properties of the stellar population.

A comparison based on observational data merely reflects the degree of agreement between different methods, but it is not possible to make an optimal choice, as the correct solution is not known, and the experiment offers limited insight about the reasons behind the observed discrepancies. The SELGIFS Data Challenge, on the other hand, provides an excellent benchmark to obtain more robust conclusions, but a significant amount of time and effort would be necessary. Let us consider, for instance, the stellar mass, as well as the mass-weighted and luminosity-weighted stellar ages and metallicities, recovered by the following methods:

\begin{itemize}
 \item Steckmap -- PSB \citep{Ocvirk06a, Ocvirk06}.
 \item Pipe3D -- SFS \citep{Sanchez16_FIT3D, Sanchez16}.
 \item Starlight -- LG \citep{Cid_Fernandes05, Cid_Fernandes13}, with the same setup as \citet{Galbany14, Galbany16_CALIFA}.
 \item Starlight -- RGB \citep[setup as in][including Salpeter\footnote{A correction factor of 0.28 dex has been applied to the stellar mass in order to match Kroupa/Chabrier normalisation. Any other effects are unaccounted for.} IMF]{Garcia-Benito17}.
\end{itemize}

The reader is referred to the aforementioned publications for a thorough description of each algorithm and the adopted parameter settings.
Here, the default choices adopted by some of the authors of the present manuscript have been used in order to provide a rough indication of the expected size of the uncertainties, but the results of this naive exercise only stress the need for much more extensive studies that rigorously explore the origin of the systematic uncertainties in each individual quantity for every different method and parameter configuration.

\begin{figure*}
\includegraphics[width=.99\textwidth]{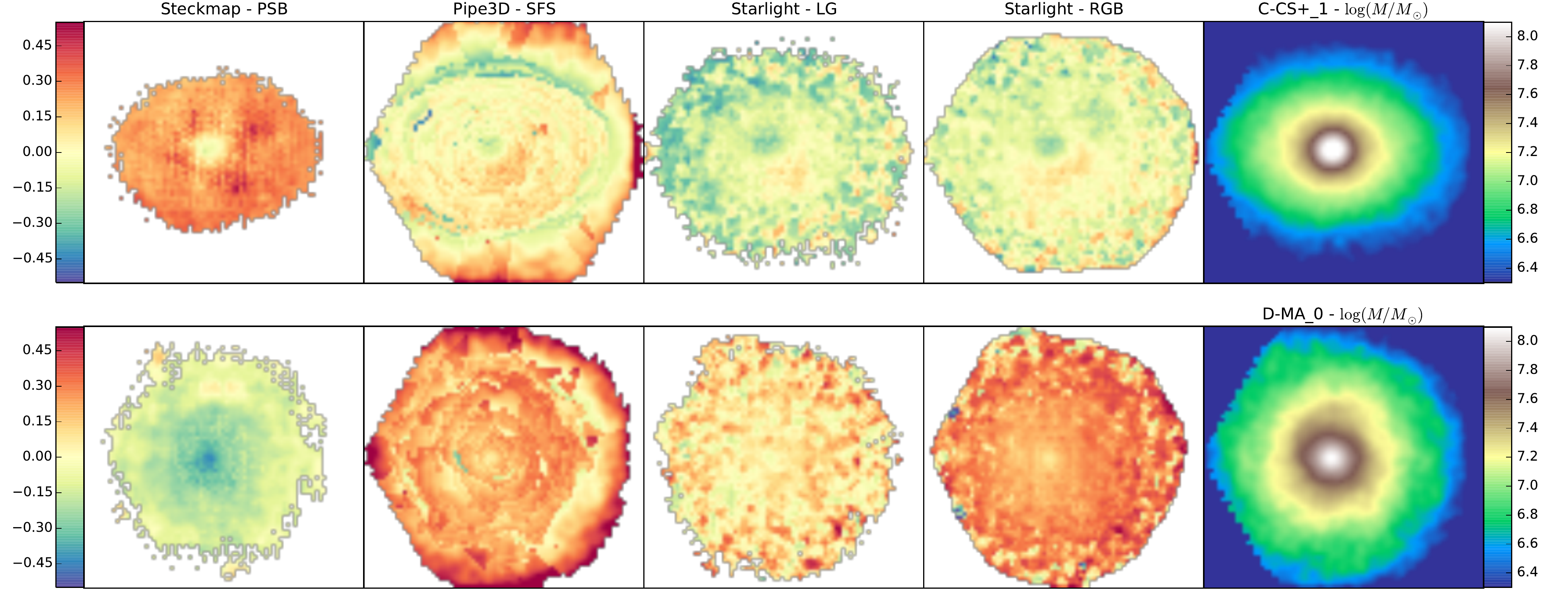}
\caption{Relative accuracy
  (measured in dex) of the stellar mass distribution recovered by different codes, compared to the true solution (shown in the rightmost column) for simulated galaxies C-CS$^+$\_1 (top) and D-MA\_0 (bottom).}
\label{fig_mass}
\end{figure*}

\begin{figure*}
\includegraphics[width=.99\textwidth]{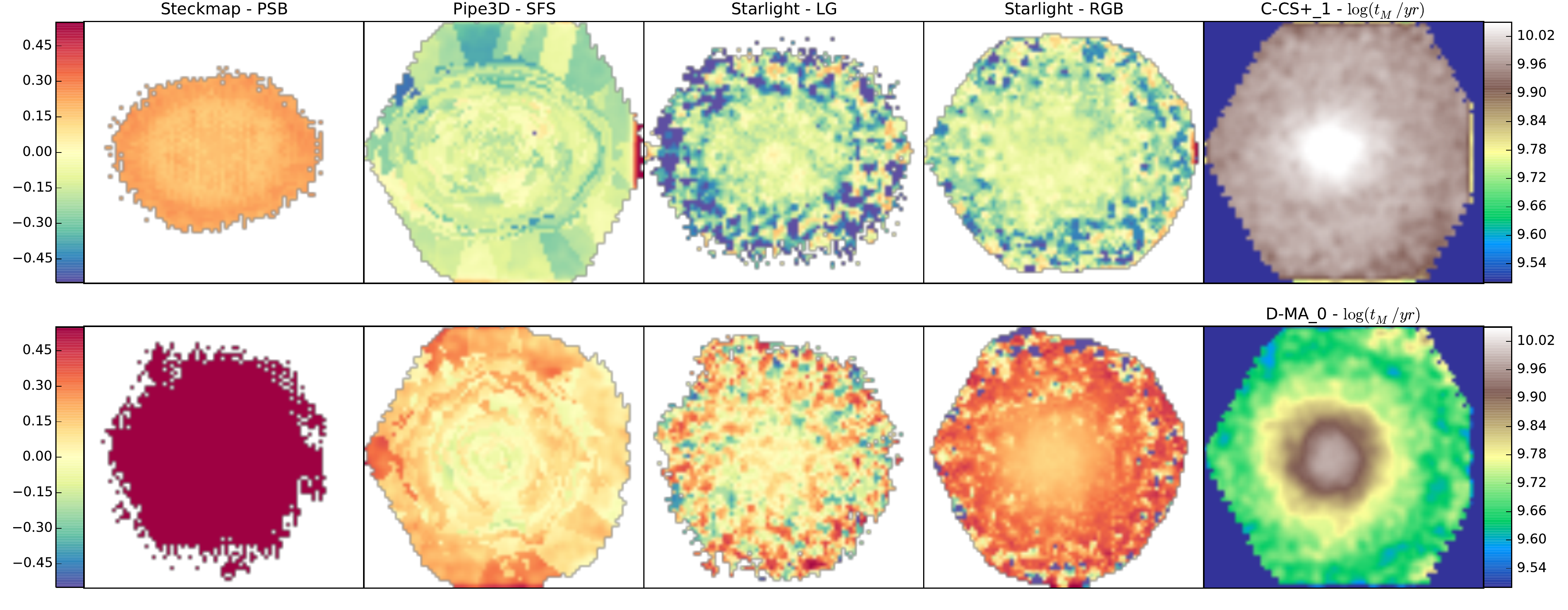}\\[5mm]
\includegraphics[width=.99\textwidth]{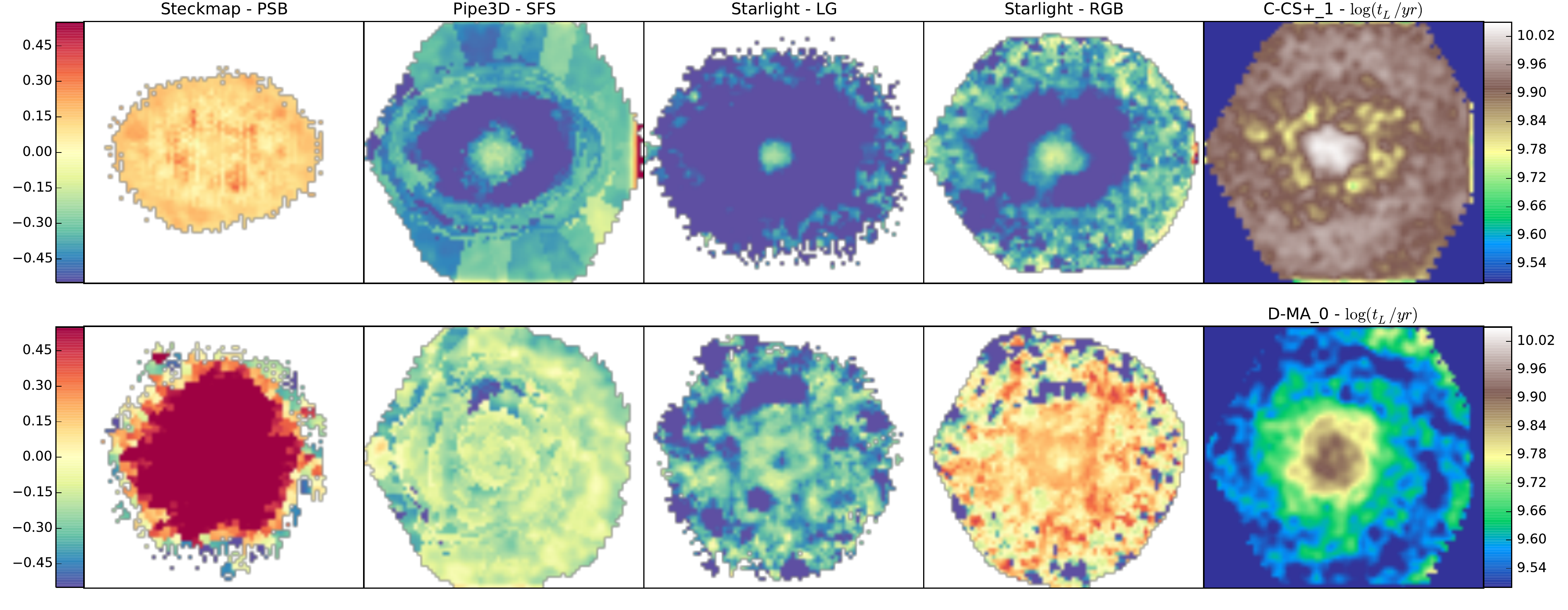}
\caption{Same as Figure~\ref{fig_mass}, for the mass-weighted (first two rows) and luminosity-weighted (last two rows) stellar age.}
\label{fig_age}
\end{figure*}

\begin{figure*}
\includegraphics[width=.99\textwidth]{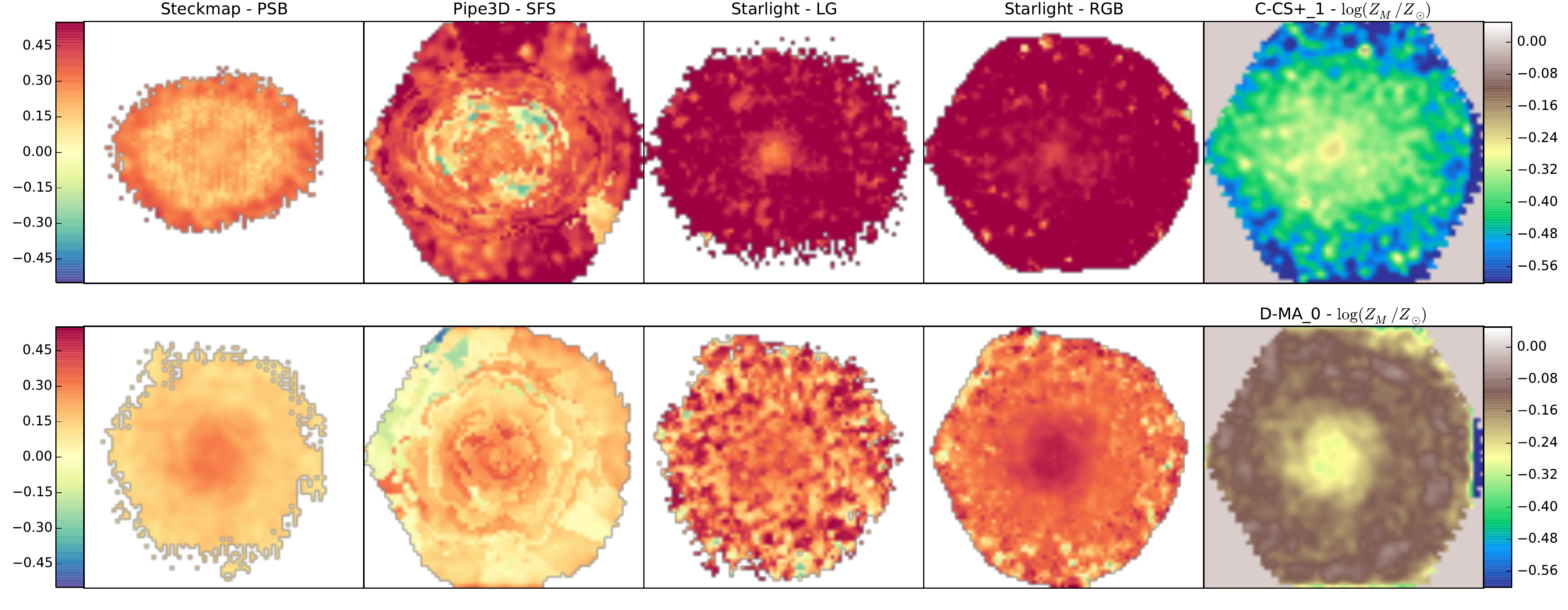}\\[5mm]
\includegraphics[width=.99\textwidth]{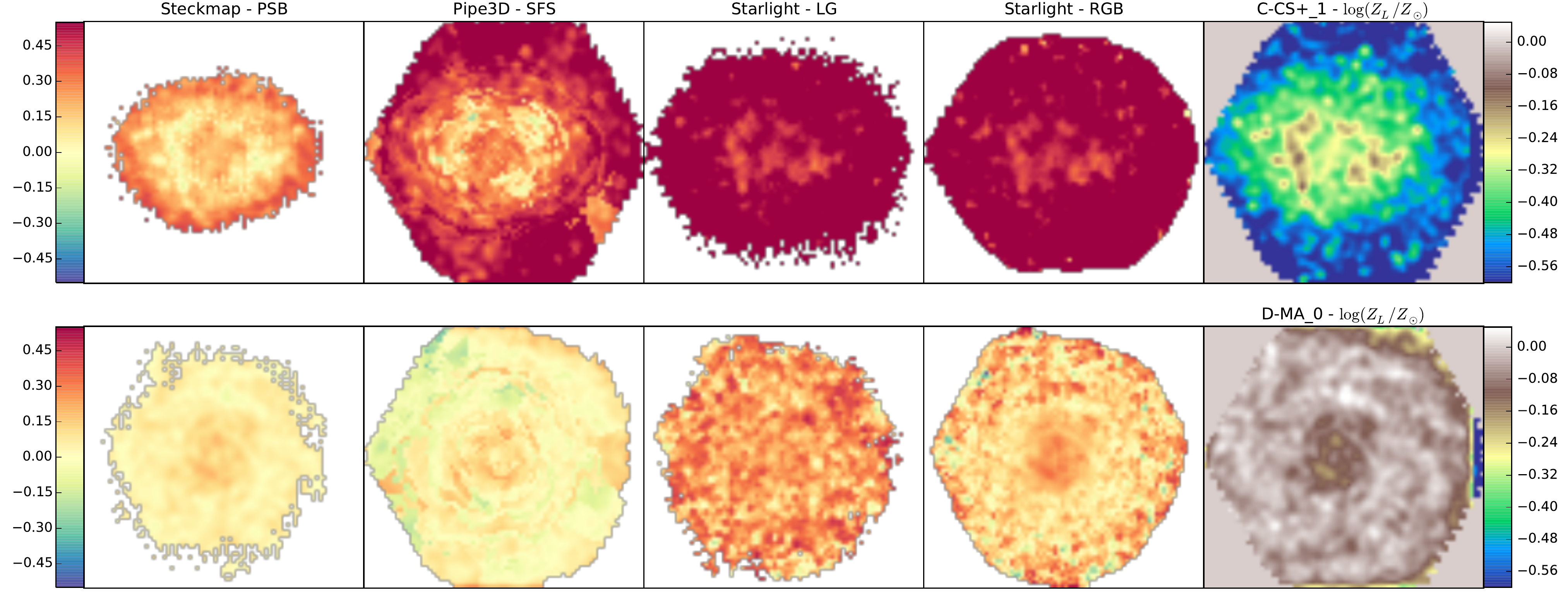}
\caption{Same as Figure~\ref{fig_mass}, for the mass-weighted (first two rows) and luminosity-weighted (last two rows) stellar metallicities.}
\label{fig_metallicity}
\end{figure*}

In essence, all the methods considered decompose the observed spectrum as a sum of simple stellar populations (SSPs) with different ages and metallicities and a set of nebular emission lines.
The average stellar properties are then estimated from the coefficients assigned to each SSP.
We compare the outputs of these methods, run on our simulated V500-like datacubes, with the true solution given by the product datacubes (convolved with the PSF) for two of our simulated galaxies (C-CS$^+$\_1 and D-MA\_0).
Results for the stellar mass, age, and metallicity are plotted in Figures~\ref{fig_mass}, \ref{fig_age}, and~\ref{fig_metallicity}, respectively.
For each quantity, we plot colour maps of the relative accuracy
\begin{equation}
 \delta_i = \log_{10}( Q_i - Q_0 )
\end{equation}
expressed in dex, where $Q_i$ denotes the value returned by each method and $Q_0$ represents the solution to be recovered.
The mass-weighted average
\begin{equation}
 \langle \delta_i \rangle_M = \frac{ \sum \mu_*(x,y)\ \delta_i(x,y) }{ \sum \mu_*(x,y) }
\end{equation}
and standard deviation
\begin{equation}
 \sqrt{ \langle \delta_i^2 \rangle_M - \langle \delta_i \rangle_M^2 }
\end{equation}
of the relative accuracy for each quantity, method, and galaxy are quoted in Table~\ref{tab:results}, where $\mu_*(x,y)$ denotes the local stellar surface density at each spaxel.

% -------------------------------------
% -------------------------------------
% -------------------------------------
% -------------------------------------
\begin{table*}
\small
\begin{center}
\begin{tabular}{ccccccc}
\hline
Galaxy & Method & mass & $\langle$ age $\rangle_M$ & $\langle$ age $\rangle_L$ & $\langle Z \rangle_M$ & $\langle Z \rangle_L$ \\
\hline
            & Steckmap  -- PSB & $~~0.249 \pm 0.071$ & $~~0.207 \pm 0.031$ & $~~0.132 \pm 0.047$ & $~~0.253 \pm 0.066$ & $~~0.208 \pm 0.106$ \\
C-CS$^+$\_1 & Pipe3D    -- SFS & $~~0.047 \pm 0.179$ & $ -0.139 \pm 0.144$ & $-0.406 \pm 0.192$ & $~~0.375 \pm 0.205$ & $~~0.438 \pm 0.206$ \\
            & Starlight -- LG  & $ -0.126 \pm 0.119$ & $ -0.237 \pm 0.210$ & $-0.717 \pm 0.279$ & $~~0.608 \pm 0.142$ & $~~0.696 \pm 0.158$ \\
            & Starlight -- RGB & $ -0.059 \pm 0.105$ & $ -0.163 \pm 0.150$ & $-0.448 \pm 0.204$ & $~~0.605 \pm 0.109$ & $~~0.657 \pm 0.133$ \\
\hline
        & Steckmap  -- PSB & $ -0.140 \pm 0.083$ & $~~7.043 \pm 0.411$ & $~~0.732 \pm 0.841$ & $~~0.166 \pm 0.048$ & $~~0.051 \pm 0.036$ \\
D-MA\_0 & Pipe3D    -- SFS & $~~0.242 \pm 0.140$ & $~~0.107 \pm 0.102$ & $ -0.185 \pm 0.099$ & $~~0.126 \pm 0.110$ & $~~0.021 \pm 0.080$ \\
        & Starlight -- LG  & $~~0.086 \pm 0.115$ & $~~0.043 \pm 0.197$ & $ -0.418 \pm 0.224$ & $~~0.310 \pm 0.137$ & $~~0.238 \pm 0.105$ \\
        & Starlight -- RGB & $~~0.248 \pm 0.135$ & $~~0.205 \pm 0.272$ & $ -0.043 \pm 0.405$ & $~~0.326 \pm 0.099$ & $~~0.138 \pm 0.113$ \\
\hline
\end{tabular}
\end{center}
\caption{Mass-weighted mean and standard deviation of the relative accuracy of each method (measured in dex) for the quantities displayed in Figures~\ref{fig_mass},~\ref{fig_age}, and~\ref{fig_metallicity}.}
\label{tab:results}
\end{table*}
% -------------------------------------

%%%%%%%%%%%%%%%%
%\end{comment}

The interpretation of these results is anything but straightforward.
With a few exceptions, all methods are able to recover the true solution within a factor of two (0.3 dex).
However, the relative accuracy depends on the quantity and method in an intricate way, that seems to be different for different galaxies.
Finding the causes of such behaviour (e.g. choice of IMF, SSP basis, fitting method, precise definition of the averaging procedure, dust extinction prescription, physical properties of the galaxy, etc.), especially in the catastrophic cases, requires a thorough analysis that is beyond the scope of the present work.

These results emphasise the influence on the adopted methodology and parameters on the recovered physical properties of the galaxies.
However, none of the prescriptions considered here can be singled out as better or worse than the others.
Even in this simple test, with very limited coverage of the huge parameter space (both on the simulation and the analysis side), it is difficult to identify clear systematic trends.
Some codes, with certain settings, are able to recover some parameters more accurately than others in one galaxy, but fare worse on another.

We argue that using any of the methods available in the literature with default parameter settings will probably incur in systematic uncertainties that are much larger than the reported statistical errors.
Understanding the effects of each parameter in detail is required in order to estimate the magnitude of the systematic uncertainty in a realistic case.
The SELGIFS Data Challenge can be extremely useful in this respect, and we encourage the reader to experiment with it.
Let us please warn, nevertheless, that the data set is not meant as a test bed to optimise the parameter settings of any given algorithm.
This would only guarantee the optimal recovery of our simulated conditions, which may not bear any close relation to reality whatsoever.

%--------------------------------------------------------------------------
\section{Summary}
\label{sec:summary}
%--------------------------------------------------------------------------

We use hydrodynamical simulations of galaxies formed in a cosmological
context to generate mock data mimicking the
Integral Field Spectroscopy (IFS) survey CALIFA \citep{Sanchez12}.
The hydrodynamical code follows, in addition to gravity and 
hydrodynamics, many other relevant galactic-scale physical processes, such as 
energy feedback and chemical enrichment 
from SNe explosions, multi-phase InterStellar Medium (ISM), 
and metal-dependent cooling of the gas.
Our hydrodynamical simulations have been post-processed with 
the radiative transfer code {\sc sunrise}, in order to obtain their 
spatially-resolved spectral energy distributions. 
These spectra contain the light emitted by the stars and the nebulae 
(young stars) in the
simulations, and include the broadening of the absorption and 
emission lines due to kinematics, as well as the extinction and 
scattering by the dust in the ISM.

The input parameters in {\sc sunrise} have been tuned to reproduce 
the properties of the CALIFA instrument in terms of 
field of view size, number of spaxels and spectral range.
After we obtain the results of the radiative transfer with {\sc sunrise}, we 
redshift the simulated spectra to match the physical size covered by 
the spaxels in the radiative transfer stage
with the angular resolution of the PMAS instrument used in CALIFA, 
and we resample and cut these spectra according to the sampling and wavelength
range of the low-resolution V500 and blue mid-resolution V1200 CALIFA setups.
We convert our spatially-resolved spectra into the V500 and V1200 data 
format of the CALIFA
DR2, and we convolve these 3-dimensional datasets with Gaussian Point 
Spread Functions 
both in the spatial and spectral dimensions, mimicking the properties 
of the CALIFA observations in terms of spatial and spectral resolution.
Finally, after we parametrize the properties of the noise in a sample of 367
galaxies both from the CALIFA V500 and V1200 datasets, we add similar 
noise to the simulated V500 and V1200 data.

Our final sample of 18 datacubes (3 objects with 3 inclinations both in 
the V500 and V1200 setups) provide observers 
with a powerful benchmark to test the accuracy and calibration of 
their analysis tools and set the basis for a reliable comparison between 
simulations 
and IFS observational data.
To this purpose we generate, together with the synthetic IFS observations, a 
corresponding set of \emph{product datacubes}, i.e. resolved maps of 
several properties computed directly from the simulations 
and/or simulated noiseless datacubes.

Although this work is specifically designed to reproduce the properties of 
the CALIFA observations, the method illustrated in this paper can 
be easily extended to mimic 
other integral field spectrographs such as MUSE \citep{Bacon04}, WEAVE 
\citep{Dalton14}, MaNGA \citep{Bundy15} or SAMI \citep{Allen15} by
changing some of the input parameters in the radiative transfer stage 
and performing a similar study of the detector noise. 
Hence, this procedure can be easily applied to generate synthetic 
observations for different IFS instruments, or for studying a  
specific science case prior to applying for observing time.
The present project can also be extended to use other hydrodynamical 
simulations, which will be very important in order to enlarge the given 
dataset and consider a more complete sample of galaxies in terms 
of morphology, total mass, stellar age and metallicity, gas content and  
merger history.

We then encourage researchers to contact the authors in case they are 
interested in obtaining simulated data mimicking the properties of different 
IFS surveys, or if they have interest in converting their hydrodynamical 
simulations into CALIFA-like datacubes.
We hope that this work would promote more collaboration and connection 
among observers and simulators, as this will be of crucial importance
in view of the several ongoing and future IFS surveys,
which will provide the community with large datasets of 
spatially-resolved properties of galaxies at different cosmic times, 
allowing to study galaxy formation physics at a higher level of detail 
than ever before.

%--------------------------------------------------------------------------
\section*{Acknowledgments}
%--------------------------------------------------------------------------

We thank the reviewer for their useful comments and suggestions,
which have helped to improve the article in several respects.
Most importantly, we are grateful for the encouragement to
implement the CALIFA dithering pattern into our pipeline.
We thank Michael Aumer for providing his simulations, 
P.-A. Poulhazan and P. Creasey for sharing the new chemical code,
and Elena Terlevich for useful comments on the manuscript. 
GG and CS acknowledge support from the Leibniz Gemeinschaft,
through SAW-Project SAW-2012-AIP-5 129,
and from the High Performance Computer in Bavaria (SuperMUC) 
through Project pr94zo. YA, CS, JC and GG acknowledge support from the 
DAAD through
the Spain-Germany Collaboration programm PPP-Spain-57050803.
The 'Study of Emission-Line Galaxies with Integral-Field Spectroscopy'
 programme (SELGIFS, FP7-PEOPLE-2013-IRSES-612701) is
funded by the Research Executive Agency (REA, EU).
YA is supported by contract RyC-2011-09461 of the \emph{Ram\'on y Cajal}
programme (Mineco, Spain).
JC has been financially supported by the ``Research Grants - Short-Term 
Grants, 2016'' (57214227) promoted by the DAAD.
JC and YA also acknowledge financial support from grant 
AYA2013-47742-C4-3-P [and AYA2016-79724-C4-1-P] (Mineco, Spain).
RGB acknowledges support from the Spanish Ministerio de Econom\'ia y
Competitividad, through projects AYA2014-57490- P and AYA2016-77846-P.
LG was supported in part by the US National Science Foundation 
under Grant AST-1311862.
PSB acknowledges support from the BASAL Center for Astrophysics and 
Associated Technologies (PFB-06).
JC would like to thank the 'Galaxy and Quasars' research group at the 
Leibniz-Institut f{\"u}r Astrophysik Potsdam (AIP) for the useful 
discussions and constructive feedback.
This study uses data provided by the Calar Alto Legacy Integral Field 
Area (CALIFA) survey (\url{http://califa.caha.es/}), 
based on observations 
collected at the Centro Astron\'omico Hispano Alem\'an (CAHA) at Calar Alto, 
operated jointly by the Max-Planck-Institut f{\"u}r Astronomie and the 
Instituto de Astrof\'isica de Andaluc\'ia (CSIC).

%--------------------------------------------------------------------------
\bibliographystyle{mnras}
\bibliography{biblio}
%--------------------------------------------------------------------------

\appendix 

%--------------------------------------------------------------------------
\section{The SELGIFS data challenge}
\label{sec:selgifs}
%--------------------------------------------------------------------------

The main goal of this work is to provide the scientific community with a 
reliable set of synthetic IFS observations, and with the corresponding
maps of directly measured properties, that allows to test existing (and 
future) dedicated analysis tools, as well as to create a benchmark for
verifying hypothesis and/or preparing observations.

The data are distributed through a web 
page\footnote{\url{http://astro.ft.uam.es/selgifs/data_challenge/}} hosted by 
the Universidad 
Aut\'{o}noma de Madrid. The description of the different 
files and their data format is presented in the following sections.

\subsection{Synthetic observations}

\label{sec:datacubes}

\begin{table*}
\begin{center}
\begin{tabular}{cccc}
\hline
HDU & Extension name & Format & Content  \\
\hline
0 & PRIMARY & 32-bit float & flux density in 
$10^{-16}$~erg~s$^{-1}$~cm$^{-2}$~\AA$^{-1}$ \\
1 & ERROR & 32-bit float & $1 \sigma$ error on the flux density \\
2 & ERRWEIGHT & 32-bit float & error weighting factor \\
3 & BADPIX & 8-bit integer & bad pixel flags (1=bad, 0=good) \\
4 & FIBCOVER & 8-bit integer & number of fibres used to fill each spaxel \\
\hline
\end{tabular}
\caption{Structure of the CALIFA FITS files in DR2 
(from \citealt{Garcia_Benito15}).}
\label{tab:data_structure}
\end{center}
\end{table*}

Our synthetic CALIFA datacubes in the two V500 and V1200 setups 
(Section~\ref{sec:califa}) are provided in 
different files, identified following the CALIFA DR2 naming convention 
GALNAME.V500.cube.fits.gz and GALNAME.V1200.cube.fits.gz 
for the V500 and V1200 respectively.
The data structure of these simulated data closely follows the one 
adopted in CALIFA, 
namely datacubes in the standard FITS file format.

The FITS header of the simulated datacubes stores only the most relevant 
keywords available in the DR2 header. Most of the DR2 keywords containing 
information about the pointing, the reduction pipeline, Galactic extinction, 
sky brightness, etc. have 
been removed.
The flux unit has been stored under 
the keyword PIPE UNITS as in the CALIFA datacubes.

Each FITS file contains the 
data for a single galaxy stored in five HDU (see 
Table~\ref{tab:data_structure}), every one of them providing different 
information according to the data format of the pipeline V1.5 used in DR2 
\citep{Garcia_Benito15}. The first two axes in the datacubes ($N_{\alpha}$, $N_{\delta}$) 
correspond to the spatial dimensions (along the right ascension and 
declination) with a $1" \times 1"$ sampling. The third dimension 
($N_{\lambda}$) represents the wavelength axis, with ranges and 
samplings described in Section~\ref{sec:califa}
and Table~\ref{tab:data_format}. 

Here we summarize the 
content of each HDU:

\noindent {\bf 0) Primary (PRIMARY)}

\noindent 
The primary HDU contains the measured flux densities in 
CALIFA units of $10^{-16}$~erg~s$^{-1}$~cm$^{-2}$~\AA$^{-1}$.\\

\noindent {\bf 1) Error (ERROR)}

\noindent 
This extension provides the values of the $1 \sigma$ noise level in each pixel,
calculated according to Eq.~\ref{eq:noise}.
In the case of bad pixels, we store a value of $10^{10}$ following the 
CALIFA data structure. \\

\noindent {\bf 2) Error weight (ERRWEIGHT)} 

\noindent
%In the CALIFA datacubes, this HDU gives the error scaling factor for 
%each pixel,
%in the case that all valid pixels of the cube are co-added; in our case we
%set all the values to 1.
{ This HDU gives the error scaling factor for each pixel, in the case
that all valid pixels of the cubes are co-added (see appendix A of
\citealt{Garcia_Benito15}).}
\\

\noindent {\bf 3) Bad pixel (BADPIX)}

\noindent
This extension stores a flag advising on potential problems in a 
pixel; in the CALIFA dataset this may occur for instance due to cosmic rays 
contamination, bad CCD columns, or the effect of vignetting. 
In our datacubes we flag as bad pixel (i.e. equal to 1) the regions in the 
spectra that are generated with the lower-resolution stellar model (see 
Section~\ref{sec:simulated_spectra}).
\\

\noindent {\bf 4) fibre coverage (FIBCOVER)}

\noindent
{ This HDU, available only from DR2, accounts for the number of fibers
used to recover the flux (see section 4.3 of \citealt{Garcia_Benito15}).}

{ In addition to these datacubes, we provide in the following files
  synthetic data free of any observational effect,
  in the same data format and physical units.
\begin{itemize}
  \item {\bf GALNAME.SED.fits:}
  { it contains the simulated SEDs prior to the addition of
    the observational effects (noise and PSFs, see
    sec.~\ref{sec:califa}).
    The
    spatially-resolved SEDs are stored in a single HDU with
    $ 78 \times 73 $ pixels in the spatial dimensions and
    1877 pixels in the wavelength dimension, with 2~\AA~ sampling
    following the V500 data format (tab.~\ref{tab:data_format}).
  }
\end{itemize}

}

%=================================================================

\begin{table}
\begin{center}
\begin{tabular}{|l|c|}
\hline
Stellar property & Units \\
\hline
Mass & $\log (\text{M}_{\odot})$ \\
Mass density & $\log (\text{M}_{\odot}/\text{pc}^2)$ \\
Mean age mass-weighted & $\log (\text{yr})$  \\
Mean metallicity mass-weighted & $\log (\text{Z}/\text{Z}_{\odot})$ \\
Mean age luminosity-weighted & $\log (\text{yr})$ \\
Mean metallicity luminosity-weighted & $\log (\text{Z}/\text{Z}_{\odot})$ \\
Mean velocity & km/s \\
Velocity dispersion & km/s \\
Star formation rate & M$_{\odot}$/yr \\
Stellar particles number & -- \\
\hline
\end{tabular}
\caption{List of the spatially-resolved stellar properties provided in 
the product datacubes.}
\label{tab:list_stellar_properties}
\end{center}
\end{table}

%======================================================================

\begin{table*}
\begin{center}
\scalebox{0.96}{
\begin{tabular}{|c|c|c|c|c|c|}
\hline
Name & Index Bandpass (\AA) & Blue continuum bandpass (\AA) & Red continuum bandpass (\AA) & Units & Reference \\
\hline
CN$_1$ & 4142.125 - 4177.125 & 4080.125 - 4117.625  &	4244.125 - 4284.125 & mag & \citet{Worthey94}\\
CN$_2$ & 4142.125 - 4177.125 & 4083.875 - 4096.375 & 4244.125 - 4284.125 & mag & \citet{Worthey94}\\
Ca4227 & 4222.250 - 4234.750 & 4211.000 - 4219.750 & 4241.000 - 4251.000 & \AA\ &\citet{Worthey94} \\
G4300 & 4281.375 - 4316.375 & 4266.375 - 4282.625 	& 4318.875 - 4335.125 &	\AA\ & \citet{Worthey94} \\
Fe4383 & 4369.125 - 4420.375 & 4359.125 - 4370.375 & 4442.875 - 4455.375 & \AA\ & \citet{Worthey94} \\
Ca4455 & 4452.125 - 4474.625 & 4445.875 - 4454.625 & 4477.125 - 4492.125 & \AA\ & \citet{Worthey94} \\
Fe4531 & 4514.250 - 4559.250 & 4504.250 - 4514.250 & 4560.500 - 4579.250 & \AA\ & \citet{Worthey94} \\
Fe4668 & 4634.000 - 4720.250 & 4611.500 - 4630.250 & 4742.750 - 4756.500 & \AA\ & \citet{Worthey94} \\
H$\beta$ & 4847.875 - 4876.625 & 4827.875 - 4847.875 & 4876.625 - 4891.625 & \AA\ & \citet{Worthey94} \\
Fe5015 & 4977.750 - 5054.000 & 4946.500 - 4977.750 & 5054.000 - 5065.250 & \AA\ & \citet{Worthey94} \\
Mg$_1$ & 5069.125 - 5134.125 & 4895.125 - 4957.625 & 5301.125 - 5366.125 & mag & \citet{Worthey94}\\
Mg$_2$ & 5154.125 - 5196.625 & 4895.125 - 4957.625 & 5301.125 - 5366.125 & mag & \citet{Worthey94}\\
Mg$b$ & 5160.125 - 5192.625 & 5142.625 - 5161.375 & 5191.375 - 5206.375 & \AA\ & \citet{Worthey94} \\
Fe5270 & 5245.650 - 5285.650 & 5233.150 - 5248.150 & 5285.650 - 5318.150 & \AA\ & \citet{Worthey94} \\
Fe5335 & 5312.125 - 5352.125 & 5304.625 - 5315.875 & 5353.375 - 5363.375 & \AA\ & \citet{Worthey94} \\
Fe5406 & 5387.500 - 5415.000 & 5376.250 - 5387.500 & 5415.000 - 5425.000 & \AA\ & \citet{Worthey94} \\
Fe5709 & 5696.625 - 5720.375 & 5672.875 - 5696.625 & 5722.875 - 5736.625 & \AA\ & \citet{Worthey94} \\
Fe5782 & 5776.625 - 5796.625 & 5765.375 - 5775.375 & 5797.875 - 5811.625 & \AA\ & \citet{Worthey94} \\
Na D & 5876.875 - 5909.375 & 5860.625 - 5875.625 & 5922.125 - 5948.125 & \AA\ & \citet{Worthey94} \\
TiO$_1$ & 5936.625 - 5994.125 & 5816.625 - 5849.125 & 6038.625 - 6103.625 & mag & \citet{Worthey94}\\
TiO$_2$ & 6189.625 - 6272.125 & 6066.625 - 6141.625 & 6372.625 - 6415.125 & mag & \citet{Worthey94}\\
H$\delta_A$ & 4083.500 - 4122.250 & 4041.600 - 4079.750 & 4128.500 - 4161.000 & \AA\ & \citet{Worthey97} \\
H$\gamma_A$ & 4319.750 - 4363.500 & 4283.500 - 4319.750 &	4367.250 - 4419.750 & \AA\ & \citet{Worthey97} \\
H$\delta_F$ & 4091.000 - 4112.250 & 4057.250 - 4088.500 & 4114.750 - 4137.250 & \AA\ & \citet{Worthey97}   \\
H$\gamma_F$ & 4331.250 - 4352.250 & 4283.500 - 4319.750 &	4354.750 - 4384.750 & \AA\ & \citet{Worthey97} \\
D4000\_n & 	& 3850.000 - 3950.000 & 4000.000 - 4100.000 & & \citet{Balogh99} \\
\hline
\end{tabular}}
\caption{List of the absorption line indices for which the strength in
each spaxel is provided, together with the definition of the 
continuum and bandpass wavelength ranges.}
\label{tab:stellar_absorption_indices}
\end{center}
\end{table*}

\begin{table}
\begin{center}
\begin{tabular}{ccc}
\hline
Species & Line centre (\AA) & Lower/ \\
 &  & upper bounds (\AA)  \\
\hline
{[Ne III]3869} & 3869.060 & $3859 - 3879$  \\
H$\delta$ & 4101.734 & $4092 - 4111$ \\
H$\gamma$ &  4340.464 & $4330 - 4350$ \\
{[O III]4363} & 4363.210 & $4350 - 4378$ \\
H$\beta$ & 4861.325 & $4851 - 4871$ \\
{[O III]4959} & 4958.911 & $4949 - 4969$ \\
{[O III]5007} & 5006.843 & $4997 - 5017$ \\
HeI 5876 &  5875.670 & $5866 - 5886$ \\
{[N II]6548} & 6548.040 & $6533 - 6553$ \\
H$\alpha$ & 6562.800 & $6553 - 6573$ \\
{[N II]6584} & 6583.460 & $6573 - 6593$ \\
{[S II]6717} & 6716.440 & $6704 - 6724$ \\
{[S II]6731} & 6730.810 & $ 6724 - 6744$ \\
\hline
\end{tabular}
\caption{List of the emission line intensities provided in the 
product datacubes. 
Line centers, lower and upper bounds are taken from the 
Sloan Digital Sky 
Survey-Garching DR7 analysis (available at the {\sc url} 
\url{http://wwwmpa.mpa-garching.mpg.de/SDSS/DR7/}).}
\label{tab:list_emission_lines}
\end{center}
\end{table}

%=====================================================================

\subsection{Product datacubes}
\label{sec:product_datacubes}

The direct calculation of the resolved (spaxel-by-spaxel) galaxy properties 
described in Sections~\ref{sec:simulations} and~\ref{sec:simulated_spectra} 
are provided in separate files. These maps have been calculated directly 
from the simulations' output, or from the noiseless synthetic spectra
prior to the addition of any observational effect.
 
The name of the files and data format are listed below:

\begin{itemize}
\item {\bf GALNAME.stellar.fits:} the file
contains the resolved maps obtained directly from the hydrodynamical 
simulations as described in Section~\ref{sec:simulations}. 
These FITS files have a single Header Data Unit (HDU) 
holding a 10-layer matrix, containing 
the nine $78 \times 73$ maps of the stellar properties in the order given
in Table~\ref{tab:list_stellar_properties}.
The header includes the information about the physical property stored 
in every layer (DESC\_*) and its units (UNITS\_*), where * refers to 
the layer number.

\item {\bf GALNAME.SFH.fits:} { it provides the resolved SFHs, storing
the mass (in solar units) formed in bins of 100 Myr. It contains a 
single HDU with a $ 78 \times 73 \times 140$ array.
The 140 time bins are ordered in lookback time, with the 
first element storing the mass formed in the last 100 Myr. 
}

\item {\bf GALNAME.Lick\_indices.fits:}
this file stores the resolved maps for the 26 Lick indices measured 
from the noiseless stellar-only datacube (see 
Section~\ref{sec:simulated_spectra}). Each file consists of a single 
HDU unit with a 26-layer matrix that contains the 
twenty-five $78 \times 73$ maps of the different absorption features listed 
in Table~\ref{tab:stellar_absorption_indices}. The header provides for 
each layer the Lick index name 
(DESC\_*) and its measured units (UNITS\_*), with * indicating the layer 
number.

\item {\bf GALNAME.nebular.fits:}
it encloses the resolved maps for the 13 nebular line intensities 
measured from the noiseless nebular-only datacube 
(Sec.~\ref{sec:simulated_spectra}). The data are stored in a single 
HDU unit with a 13-layer matrix, containing all the thirteen 
$78 \times 73$ maps of the nebular lines given in 
Table~\ref{tab:list_emission_lines}. The header stores the line names 
(DESC\_*), rest frame
wavelengths (LAMBDA\_*) and units (UNITS\_*) for each layer * in
the file.

\end{itemize}

In order to provide results directly comparable with the ones generated
by the observational algorithms applied to the synthetic datacubes, 
maps at the same spatial resolution of the synthetic datacubes
are additionally available. 
These have been obtained convolving the stellar maps with a 2.5 
FWHM Gaussian kernel, and the 
synthetic spectra with a 2.5'' FWHM PSF before extracting the Lick 
indices and the nebular line intensities as described in 
Section~\ref{sec:simulated_spectra}. 
Notice that  when we compute the logarithmic quantities in the stellar maps
the PSF is added prior to the calculation of the logarithm.

%====================================================================

\bsp

\label{lastpage}

\end{document}